\documentclass[a4paper]{cas-dc}

\usepackage[authoryear]{natbib}

\newcommand{\chandra}{{\textit{Chandra}}}
\newcommand{\rxte}{{\it RXTE}}
\newcommand{\asca}{{\it ASCA}}
\newcommand{\rosat}{{\it ROSAT}}

\newcommand{\suzaku}{{\it Suzaku}}
\newcommand{\xmm}{{\it XMM-Newton}}

\newcommand{\swi}{{\it Swift}}

\newcommand{\nustar}{\textit{NuSTAR}}

\newcommand{\lumcgs}{ergs~s$^{-1}$}

\def\tsc#1{\csdef{#1}{\textsc{\lowercase{#1}}\xspace}}
\tsc{WGM}
\tsc{QE}
\tsc{EP}
\tsc{PMS}
\tsc{BEC}
\tsc{DE}

\def\aj{AJ}
\def\apj{ApJ}

\def\aap{A\&A}
\def\mnras{MNRAS}
\def\apjs{ApJS}

\def\nat{Nature}
\def\memsai{Mem. SAI}

\def\aaps{A\&AS}
\def\pasp{PASP}
\def\pasj{PASJ}
\def\nar{NewAR}
\def\araa{ARA\&A}
\def\actaa{Ac. Ast.}
\def\aapr{A\&A Rev.}

\begin{document}
\let\WriteBookmarks\relax
\def\floatpagepagefraction{1}
\def\textpagefraction{.001}
\shorttitle{Accretion flows of AWBs  in the X-rays}
\shortauthors{\c{S}. Balman}

\title [mode = title]{Accretion Flows in Nonmagnetic White Dwarf Binaries as Observed in X-rays}                      



\author[1]{\c{S}\"olen Balman}[type=,
                        auid=,bioid=1,
                        prefix=,
                        role=,
                        orcid=0000-0001-6135-1144]
\cormark[1]
\ead{solen.balman@gmail.com; solen.balman@khas.edu.tr}


\address[1]{Kadir Has University, Faculty of Engineering and Natural Sciences, Cibali 34083, Istanbul, Turkey}







\cortext[cor1]{Principal corresponding author}



\begin{abstract}

Cataclysmic Variables (CVs) are compact binaries with white dwarf (WD)
   primaries. CVs and other accreting WD binaries (AWBs) are useful laboratories for studying accretion flows, gas
   dynamics, outflows, transient outbursts, and explosive nuclear
   burning under different astrophysical plasma conditions. They have been studied over
   decades and are important for population studies of galactic X-ray
   sources. Recent space- and ground-based high resolution
   spectral and timing studies, along with recent surveys indicate that
   we still have observational and theoretical complexities yet to answer. 
   I review accretion in nonmagnetic AWBs in the light of X-ray observations. 
   I  present X-ray diagnostics of accretion in dwarf novae and the disk outbursts, the nova-like systems, and the
   state of the research on the disk winds and outflows in the nonmagnetic CVs together with comparisons and relations to classical and recurrent nova systems, AM CVns and Symbiotic systems. I discuss how the advective
   hot accretion flows (ADAF-like) in the inner regions  of accretion disks (merged with boundary layer zones) in nonmagnetic CVs explain most of the discrepancies and complexities that have been encountered
   in the X-ray observations.  I stress how flickering variability studies from optical to
   X-rays can be probes to determine accretion history and  disk structure  together with how the
   temporal and spectral variability of CVs are related to that of LMXBs and AGNs. Finally, I discuss the nature of accretion in nonmagnetic WDs in terms of ADAF-like accretion flows, and elaborate on the solutions it brings
   and its complications, constructing an observational framework to motivate new theoretical calculations that introduce this flow-type in disks, outflow and wind models together with disk-instability models of outbursts and nova outbursts in AWBs  and  WD physics, in general.

\end{abstract}


\maketitle

\section{Introduction}

Cataclysmic Variables (CVs) and related systems (e.g., AM CVns, Symbiotics) are referred as
accreting white dwarf binaries (AWBs). They are compact binary systems with white dwarf (WD) primaries.
They constitute laboratories to study accretion flows,
gas dynamics, outflows, transient outbursts, and explosive nuclear
burning. In most CVs, accretion is via Roche Lobe overflow and disk accretion.
The donor star is a late-type main sequence star or sometimes a slightly evolved star. Systems show
orbital periods of 1.4-13 hrs with few exceptions out to 2-2.5 day binaries.
AM CVn stars also display Roche lobe overflow with the possibility of stream impact accretion
leading to hot spots on the WDs. 
These systems host either two WDs (double-degenerate systems)
or a He-star plus a WD binary.  AM CVns are ultra compact systems with binary periods between 5 and 65 min.
Another class of AWBs are Symbiotics. The accretion in these systems is sustained by winds, not in general  a disk. 
However, there are indications of temporary disk formation or existing disks in some systems.  
Donors in Symbiotics are giants (e.g., mira variables, red giants). Symbiotic systems have 
orbital periods on the order of several 100 days  to several 100 years. Note that there can be exceptions that may be classified as both CV 
and symbiotic star where orbital periods can be several days (e.g., 4-6 days).

CVs have two main catagories \citep{1995Warner}. An accretion
disk forms and reaches all the way to the WD in cases where the magnetic field of the WD is weak or nonexistent ( $B$ $<$ 0.01
MG), such systems are referred as nonmagnetic CVs and are 
characterized by their eruptive behavior \citep[see][]{1995Warner,2012Balman-mem,2017Mukai}.
The other class is the
magnetic CVs (MCVs), divided into two sub-classes according to the
degree of synchronization of the binary. Polars have strong magnetic fields in the range of 20-230 MG,
which cause the accretion flow to directly channel onto the magnetic pole/s of the WD inhibiting the formation of an accretion disk.
The magnetic and tidal torques cause the WD rotation to synchronize with the binary orbit. Intermediate Polars, which
have a weaker field strength of 1-20 MG, are asynchronous systems
\citep[see][]{1995Warner,2012Mouchet,2017Mukai}.

This review will mainly include nonmagnetic AWBs. A review on X-ray observations of MCVs by de Martino et al. can be found in 
this special issue, and see also \citet{2019Suleimanov} for joint theoretical and observational facts on  X-ray emission of MCVs. In addition, 
a small subclass of nonmagnetic CVs known as Super Soft X-ray sources (SSS), where the WD exhibits steady nuclear burning of H, is not
particularly discussed since a review can found in \citet{2010Kahabka,2010Charles} and partly discussed in the papers
contributed by Page et al., Ness J., and Orio, M. in this special issue.

\section{Nonmagnetic cataclysmic variables}

In nonmagnetic CVs the transferred matter
forms an accretion disk that reaches all the way to the WD.
Standard accretion disk theory \citep{1973Shakura} predicts half of the accretion
luminosity to originate
from the disk and the other half to emerge from the
boundary layer (BL) \citep{1974Lynden-Bell,1981Pringle}. During low-mass accretion rates, $\dot M_{acc}$$<$10$^{-(9-9.5)}$M$_{\odot}$ yr$^{-1}$, in this prescription, 
the BL is optically
thin \citep{1993Narayan,1999Popham} and emits mostly in the hard
X-rays (kT$\sim$10$^{(7.5-8.5)}$ K).
According to standard accretion disk theory, for higher accretion rates, $\dot M_{acc}$$\ge$10$^{-(9-9.5)}$M$_{\odot}$ yr$^{-1}$, the BL is expected to be
optically thick \citep{1995Popham,1995Godon,2013Hertfelder,2014Suleimanov,2015Hertfelder}
and emits in the soft X-rays and EUV (kT$\sim$10$^{(5-5.6)}$ K). 
\citet{1993Narayan} show that the optically thin BLs
can be radially extended, advect part of the
energy to the WD as a result of their inability to cool. 
Observations in the X-rays indicate that almost all systems in quiescence and outburst show an
optically thin hard X-ray emitting component revealing the nature of the BLs (more discussion will be presented later). 
The standard disk is often found inadequate to model disk-dominated, high state CVs in the UV, as well as some eclipsing quiescent dwarf nova and generates a spectrum that is bluer than the observed UV 
spectra indicating that the expected hot optically thick inner flow of the BL is not existent
 \citep[][and references therein]{1990Wood,2004Baptista,2005Linnell,2010Linnell,2007Puebla,2017Godon}.  As a result, a recent disk model of high state CVs has used a truncated inner disk (also accounts for the quiescent CVs). 
Instead of removing the inner disk, the authors impose a no-shear boundary condition at the truncation radius, lowering the disk temperature 
and generating a spectrum that yields better fits to the UV data successfully \citep[][and references therein]{2017Godon}. This allows for optically thin slightly extended BLs in the high 
states of CVs. However, the extent and nature of the X-ray flows are not fully justified.  

Dwarf novae (DNe) are a class of nonmagnetic CVs where matter is transferred by an accretion disk
at a low rate in quiescence, $\le$10$^{-10}$M$_{\odot}$ yr$^{-1}$.  Every few weeks to months or sometimes with longer durations, 
intense accretion (outburst) of days to weeks is observed where $\dot{\rm M}$ increases to a high state 
as a result of an instability in the accretion disk
\citep{1995Warner,2004Mauche}. The total disk energy involved in the outburst
of brightness is 10$^{39}$-10$^{40}$ erg where the brightening changes in a range $\Delta$m=2-6 in magnitude.
The DNe are divided into three major subclasses. U Gem types have orbital periods over 3 hrs, showing
no superoutbursts and more rare outbursts. Z Cam subtypes have occasional standstills between the outbursts with brightness that 
does not go back to the original quiescence level for a prolonged time after the outburst.
SU UMa subclass have orbital periods below 2 hrs. They have definite normal outbursts and superoutbursts which are 
longer and brighter than normal outbursts
(followed by superhumps that are a periodic brightness variation of an eccentric and precessing CV accretion disk, with a period within a few percent of the orbital period of the system).

The nonmagnetic nova-likes (NLs) are found mostly in a state of high mass accretion rate with  a
few $\times$10$^{-9}$M$_{\odot}$ yr$^{-1}$ to a few $\times$10$^{-8}$M$_{\odot}$ yr$^{-1}$. They
have winds that are about or less than 1\% of the mass accretion rate,
with velocities 200-5000 km/s \citep[see][and references therein]{2004Kafka,2002Long,2014Balman-apj,2017Godon}.
The VY Scl-type subclass exhibits
high states and occasional low states of optical brightness while the UX UMa sub-type
remains in the high state \citep{1995Warner}.  All NLs show emission lines while UX UMa stars
also exhibit broad absorption lines at optical or UV wavelengths. 

The last class of nonmagnetic CVs are the classical and recurrent novae. They are the third most violent explosions associated with a star
after gamma-ray bursts and supernovae (E$_{Total}$$\simeq$ 10$^{43}$-10$^{46}$ erg). 
They are not due to outbursts of accretion in the disks as a result of
thermal disk instabilities but are due to explosive ignition of accreted H matter on the surface of the WDs as the stable critical
pressure is surpassed \citep[see][and references therein]{2008Bode}.

The space density of nonmagnetic CVs is 10$^{-5}$-10$^{-4}$ pc$^{-3}$ as theoretically calculated \citep{1992deKool,1993Kolb}.
However, the nonmagnetic CVs have about (0.2-1)$\times$10$^{-5}$ pc$^{-3}$ density as calculated using
X-ray luminosity function of the Galaxy  \citep[see][and references therein]{2015Pretorius}. A more up-to-date value is 
(0.4-5.4)$\times$10$^{-5}$ pc$^{-3}$ using the \textit{Gaia} DR2 database \citep[][see also Pala et al. in this special issue]{2019Pala}.
The majority of  progenitors of CVs (Pre-CVs) have low-mass WDs  with an average of 0.5 M$_{\odot}$  \citep{1993Kolb,1996Politano}. 
On the other hand, observations of  WDs in CVs indicate that the average mass is larger than 0.8 M$_{\odot}$
 \citep{2011Zorotovic, 2016Nelemans, 2016Schreiber} which is supportive of additional mechanisms of angular momentum loss 
 that have been recently suggested  in the latest evolutionary models. See also the review by Zortovic and Schreiber in this special issue.
The period minimum in the CV distribution is 65 min as calculated theoretically  \citep{1999Kolb,2001Howell}, however
observations indicate a minimum period of 79.6$\pm$0.2 min \citep[][and references therein]{2019McAllister}.  
The discrepancies in the observed and theoretical, minimum period and space density are still open problems in the current understanding of the CV evolution together with
the WD mass problem. 

\subsection{X-ray observations of dwarf novae in quiescence and outburst}

The quiescent X-ray spectra  of DN are mainly characterized by a multi-temperature isobaric cooling flow
model of plasma emission (i.e., a collisionally ionized plasma in equilibrium)  at T$_{max}$=6-55 keV with accretion rates of
10$^{-12}$-10$^{-10}$ M$_{\odot}$ yr$^{-1}$ which indicate optically thin hard X-ray emitting BLs in low accretion rate states.
The X-ray line spectroscopy indicates narrow emission lines (brightest OVIII K$\alpha$)
and near solar abundances, with a 6.4 keV iron resonance line expected to be due to reflection from the surface of the WD.
The detected Doppler broadening in lines during quiescence is $<$750 km s$^{-1}$ at sub-Keplerian velocities (i.e., expected large Doppler broadening due to fast rotation in the standard 
BLs is not seen) with electron densities $>$10$^{12}$ cm$^{-3}$
\citep[see][]{2005Baskill,2006Kuulkers,2006Rana,2005Pandel,2011Balman,2012Balman-mem,2017Wada}.
Figure 1 shows some typical quiescent DNe spectra obtained using  \xmm\ data.
The total X-ray luminosity during quiescence is 10$^{28}$ -10$^{32}$ erg s$^{-1}$.
A lack of BL emission in the X-rays have been suggested due to the low L$_x$/L$_{disk}$ ratio \citep[see][]{2006Kuulkers}. Although
the standard steady-state disk accretion requires that L$_x$ and L$_{disk}$ be roughly  the same, the observations do not support this 
scenario and instead find the ratio to be between 0.1-0.0001 \citep[see][and references therein]{2006Kuulkers}.
It has been suggested for quiescent DN that if the WD emission is removed carefully and
disk truncation is allowed to some extent, this ratio is $\sim$1 \citep{2005Pandel}, however cases where this ratio is $\sim$ 0.05 exist 
(ratio has been calculated from accretion rate for a low inclination DN; Nabizadeh \& Balman 2020 see this special issue).

DNe outbursts are brightennings of the accretion disks as a result of thermal-viscous instabilities as summarized
in the Disk Instability Model \citep[DIM][]{2001Lasota,2004Lasota}. A new DIM version using viscosity and disk vertical structures obtained through MRI 
(Magneto-resonance instability; Balbus \& Hawley 1991) simulation has been recently developed by \citet{2016Coleman}. Some recent disk and DIM calculations including irradiation of the secondary
and the disk, and disk truncation  (magnetic in this case) can accomodate DIM characteristics in the 
CV light curves and the hysteresis effect better \citep{2018Dubus,2017Hameury,2020Hameury}. Hysteresis effect (dependence of the state of a system on its accretion history) may be similar to the X-ray Binaries (XRBs) if UV emission is also included in the observational comparisons and calculations together with X-rays (which, then, it does not properly reflect the isolated BL emission).
See, also, the review by J-M. Hameury  on accretion and disk instabilities in this special issue.

\begin{figure}
\label{fig:1}
\includegraphics[width=8.4cm,height=14cm,angle=0]{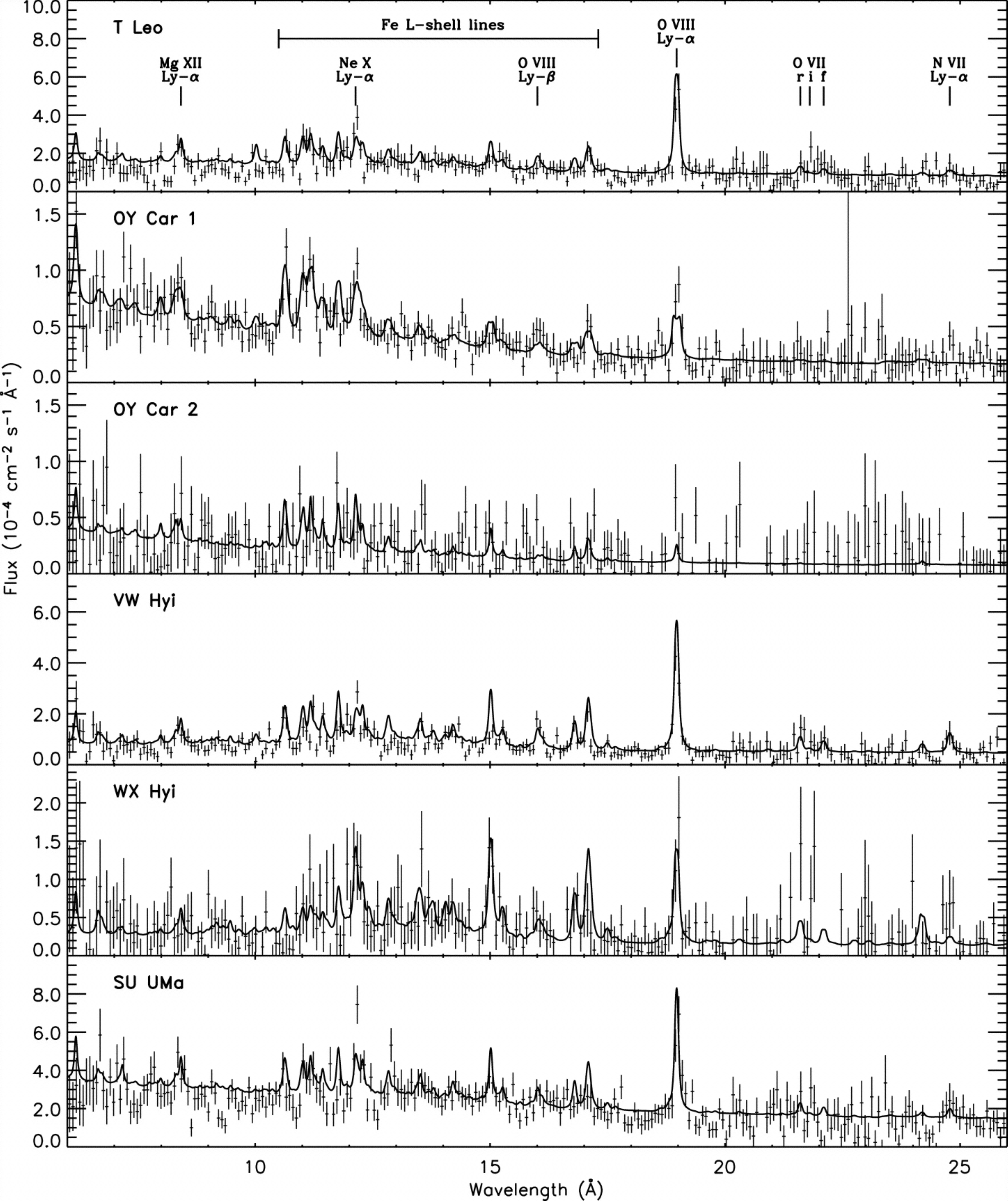}
\includegraphics[width=8.4cm,height=5cm,angle=0]{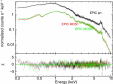}
\caption{Sample of DN high resolution \xmm\ RGS spectra between 0.4-2.5 keV \citep{2005Pandel}. At the bottom is the low resolution \xmm\ EPIC spectra of RU Peg between 0.3-10.0 keV \citep{2011Balman}.  
}
\end{figure}

DNe X-ray spectra during outburst differ from those during  quiescence because the
accretion rates are about 100 times greater \citep[][]{2011Knigge}, where the BL is expected to be optically thick and emit  in the 
extreme ultraviolet (EUV) and soft X-rays (see sec. 2). On the other hand,
soft X-ray/EUV emission and temperatures in a range 5-25 eV are detected from only around a handfull of systems (5-6)
\citep[e.g.,][]{1995Mauche,2000Mauche,1996Long,2009Byckling}. The absence of the soft components are not due to absorption since 
most DN have low interstellar extinction with hydrogen column density N$_H$$<$6$\times$10$^{20}$ cm$^{-2}$ \citep[][and references therein]{2006Kuulkers,2011Patterson,2017Godon}. 
Note that these systems do not expel matter  (e.g., as in novae) in the outburst, but only some wind/outflows (see sec. 4).
As a second and more dominantly detected emission
component (in every outburst), DN show hard X-ray emission during the outburst stage however, at a lower flux level and X-ray
temperature compared with the quiescence all throughout the outburst \citep[e.g.,][]{2003Wheatley,2004McGowan,2009Ishida,2010Collins,2011Fertig,2015Balman}. This
hard X-ray component shows evolution during the outburst into the quiescence (where the soft X-ray component is not seen to evolve when it is on).
For comparison, Figure 2 shows the hard X-ray fluxes, the X-ray temperatures (\rxte), and the optical light curves 
(AAVSO) of SS Cyg in quiescence and outburst. This source shows the soft X-ray emission component (not indicated in Figure 2) which yields an increased level 
of X-ray emission in total, during the outburst. 
Only few other DN show increased level of X-ray emission \citep{2009Byckling,2006Guver,2016Semena} mostly as a result of the detection of the
soft X-ray component (except U Gem). 
The total X-ray luminosity during the outburst is typically in the range 10$^{29}$- a few$\times$10$^{32}$ erg s$^{-1}$. An exception is SS Cyg, 
which has a luminosity about 50 times greater. The grating spectroscopy (using \xmm\  and \chandra) of the outburst 
data indicates  large widths for lines with velocities in excess of 1000 km s$^{-1}$
mostly of H and He-like emission lines (C,N,O,Ne, Mg, Si, Fe, ect.) \citep{2004Mauche,2005Pandel,2006Rana,2006Guver,2008Okada}.

A characteristic of some DN outburst light curves are the
UV and X-ray delays in rise to outburst (w.r.t. optical) indicating optically thick disk truncation (see also sec. 5) \citep[][and references therein]{1994Meyer,1999Stehle}.
These delays are a matter of several hours (up to a day) that need dedicated simultaneous
multi-wavelength observations.
No  eclipses or  distinct orbital variations are seen during outbursts (particularly of the soft X-ray emission)
 \citep[e.g.,][]{1999Pratt,2009Byckling} indicating that the extent of  the X-ray emitting region is radially extended and/or vertically high.
On the other hand, there has been few eclipse detections (e.g., Z Cha, HT Cas) during  quiescence in the X-rays with one in the high state CV, UX UMa. Detailed analyses
indicate that these are nondetections rather than flat-bottomed very low count rate regions of the light curves. Since these sources are in general, dim in the X-rays
with low count rates,  the local absorption and transparency effects can create episodic nondetections in the light curves. These nondetections are more likely 
dips or low transparency regions on the (e.g., outer) disk as in dipping low-mass X-ray binaries (LMXBs) where such transparency effects will follow the size of the absorbing region 
which can be similar to the size of the WD and create confusion.
Both in SS Cyg \citep{2004McGowan} and in SU UMa \citep{2010Collins}, 
the X-ray flux between outbursts have been found to decrease which is contrary to the expectations of the DIM model (which can be explained via disk truncation as the inner disk pulls out in quiescence).

\begin{figure}
\label{fig:2}
\includegraphics[width=8.4cm,height=6.2cm,angle=0]{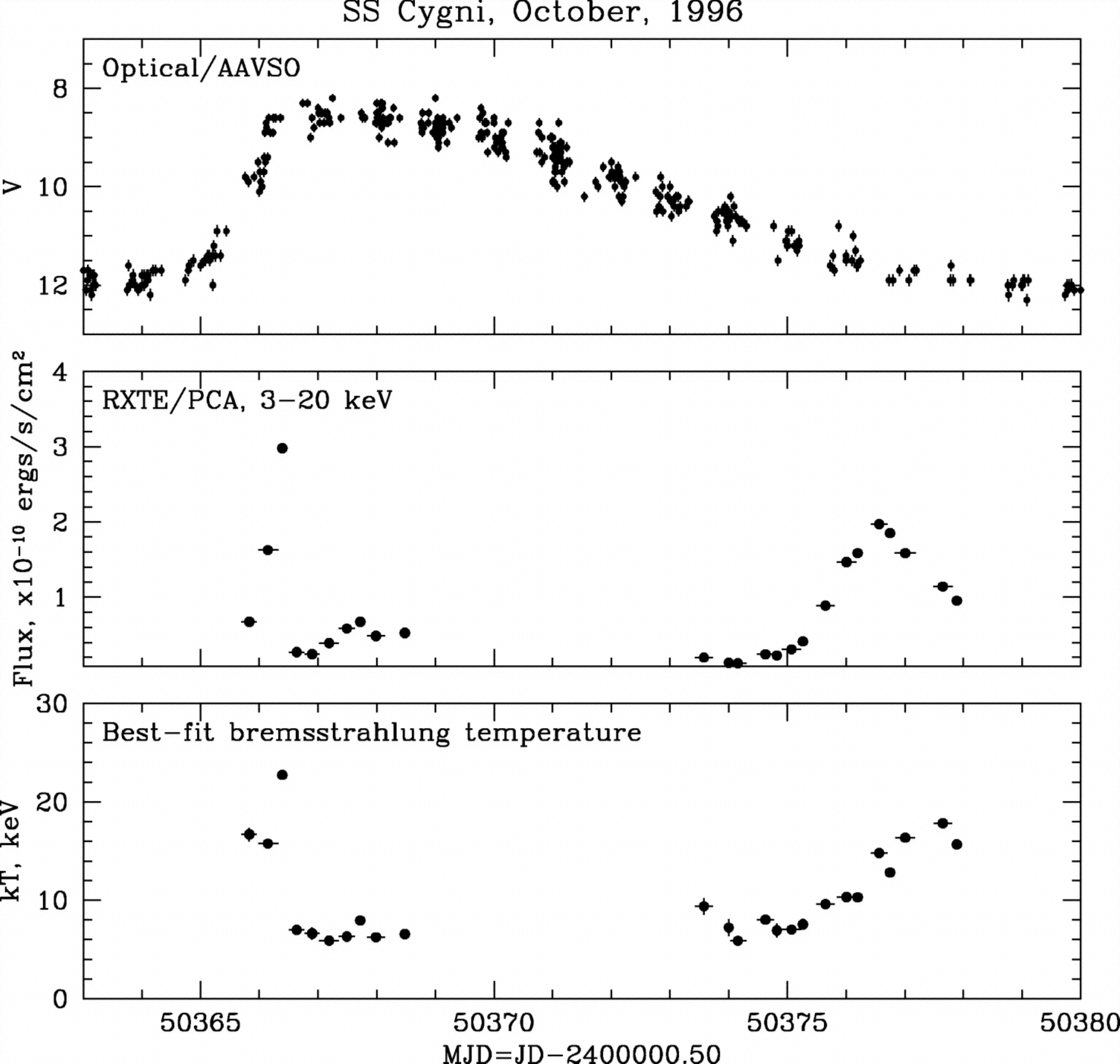}
\includegraphics[width=8.4cm,height=6.2cm,angle=0]{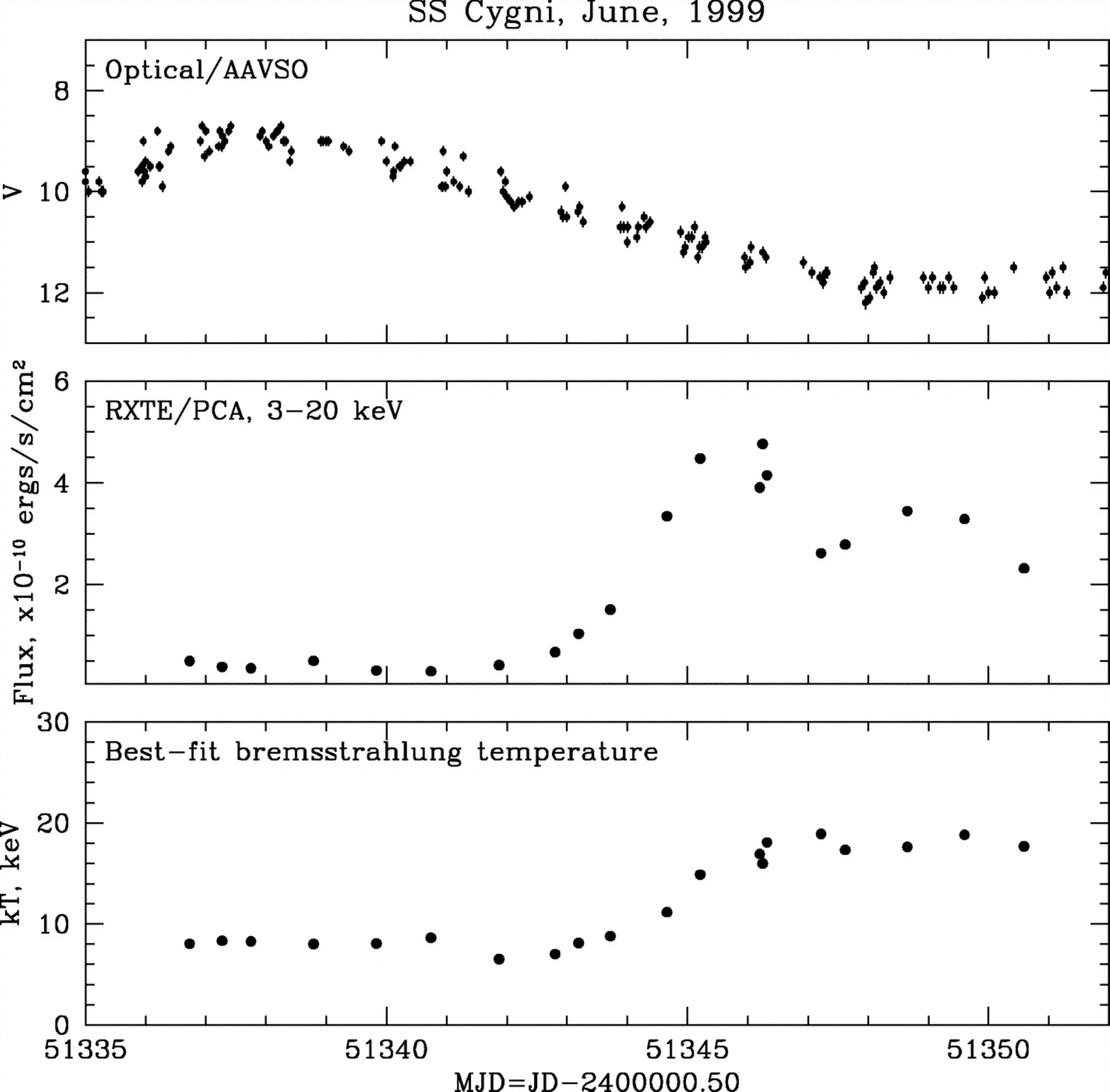}
\includegraphics[width=8.4cm,height=6.2cm,angle=0]{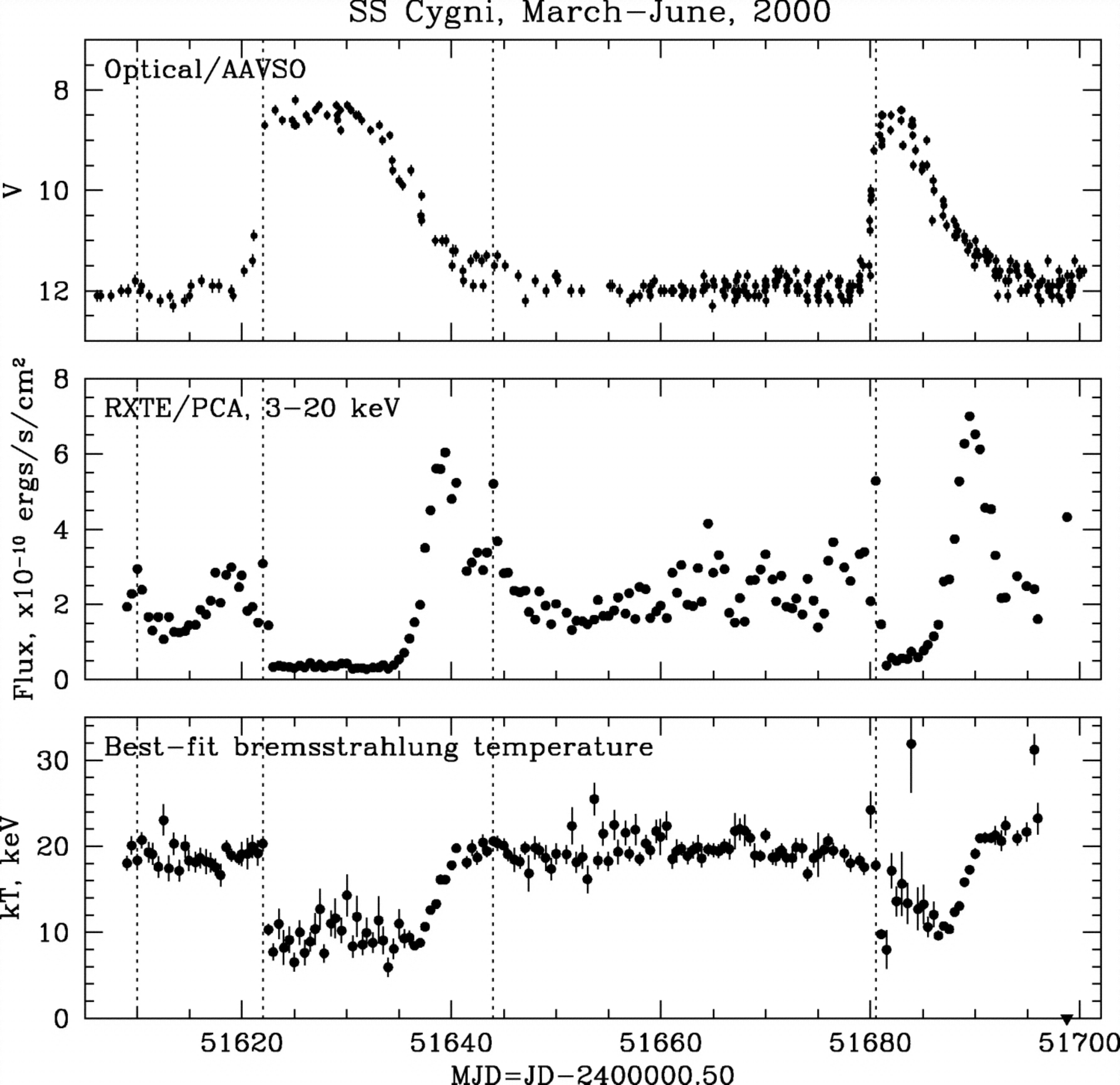}
\caption{The X-ray (RXTE) and optical (AAVSO) light curves and corresponding X-ray fluxes and Bremsstrahlung temperatures for
outbursts and following quiescent phases of SS Cyg. It is obtained from \citet{2004McGowan}. 
}
\end{figure}

\subsection{X-ray observations of nova-likes}

Observations of nonmagnetic CVs at low mass accretion rates
($\dot M_{acc}$$\le$10$^{-10}$ M$_{\odot}$ yr$^{-1}$)
(DN in quiescence),
have yielded quiescent hard X-ray spectra consistent with an optically thin
multi-temperature isobaric cooling flow model of plasma emission as described in sec.(3). 
At high mass accretion rates ($\dot M_{acc}$$\ge$10$^{-9}$ M$_{\odot}$ yr$^{-1}$),
as opposed to standard steady-state disk model calculations (where soft X-ray emission is expected from BLs) , observations of nonmagnetic CVs
(namely NLs) show a hot optically thin X-ray
source as found in all observations with luminosities $\le$ a few $\times$10$^{32}$ \lumcgs\
\citep{1985Patterson,1996vanTeeseling,1995Schlegel,1998Greiner}.
Later, some NLs were studied with \asca, \xmm, \chandra\ and \swi\
yielding spectra consistent with double MEKAL models or multi-temperature plasma models
with luminosities $\le$ a few $\times$10$^{32}$ \lumcgs
\citep[e.g.,][]{2002Mauche,2004Pratt,2014Page-nl,2014Balman-apj,2014Zemko,2017Dobrotka}.
Note that the absence of the soft X-ray emission is not due to absorption and 
most NLs have low interstellar and intrinsic absorption with hydrogen column density N$_H$$\le$6$\times$10$^{20}$ cm$^{-2}$ with very few that goes to 2-4 times this value \citep[][and references therein]{2006Kuulkers,2017Godon}.
A study by \citet{2014Page-nl} on the VY Scl-type NL V751 Cyg indicates, yet again an optically thin plasma emission spectrum in the optically high state, with additional
absorption effects in the X-ray spectrum (i.e., existence of a possible warm absorber) on the orbital plane along with less variability of the X-ray emission as opposed to the
UV emission (in magnitude). \citet{1999Greiner} claims that this source has a 15 eV blackbody-like super soft X-ray emission at an occasional optical low state, however this claim has not been confirmed. 
Besides, the spectral and timing analysis of the \xmm\ data of MV Lyr  (VY Scl-type) suggest a geometrically thick corona that surrounds an inner geometrically thin disk \citep{2017Dobrotka}. 
In contrast to this study, MV Lyr has also been found to show optical bursts interpreted as the presence of unstable, magnetically regulated accretion (magnetically gated accretion)  \citep{2017Scaringi}.  A range 
of 20-100 kG magnetic field is needed to build up material around the magnetospheric boundary which accretes onto the white dwarf, producing bursts.
However, note that the frequency breaks of 1 mHz detected by \citet{2012Scaringi} in the power spectra for the optical data of MV Lyr yield too high a magnetic field value inconsistent with the 
magnetically gated accretion scenario or any magnetic CV approximation, see  sec. (3).

\subsection{Advective Hot Flows (ADAF-like) as the Origin of X-ray Emission in CVs}

As summarized in the previous sub-sections the general characteristics of X-ray emission in CVs of particularly high states and low/quiescent states, are 
not as predicted from the standard accretion flows in steady-state disks and the evolution in thermal-viscous instability driven disk outbursts. Thus, the
standard disk theory and BL formation, do not explain the detected X-ray observations, properly. 
 
Using the present X-ray telescopes with higher sensitivity,
and better spectral resolution, some NLs previously studied with \rosat; namely,  
the VY Scl-type CVs, BZ Cam and MV Lyr, and the UX UMa-type CV, V592 Cas  were investigated with  
\swi\ for a better understanding of their X-ray characteristics \citep[see][for details]{2014Balman-apj}.
This detailed study indicates that spectra of these sources are consistent with a multi-temperature plasma
emission where the X-ray temperatures are in a range kT$_{max}$=(21-50) keV 
and the X-ray emitting
plasma is virialized. \swi\ does not detect the 6-7 keV Fe emission lines, but no other NL has been found to show significant iron line complex
in this band using \swi. \citet{2014Balman-apj} does not detect any soft X-ray emission component and calculate 7 eV as an upper limit for any soft X-ray emission using
 the \rosat\ data of these three NLs that is consistent with WD temperatures.
The ratio (L$_{x}$/L$_{disk}$) (L$_{disk}$ from the UV-optical wavelengths)
yields considerable inefficiency in the optically thin BL/X-ray emitting region by $\sim$ 0.01-0.001. Note that this is in agreement with the previous ratios given in section (2.1).
Moreover, the power-law indices of the
temperature distribution of the plasma show departures from the isobaric cooling-flow-type plasma in equilibrium (i.e., models used to explain quiescent DNe X-ray emission).
The authors also suggest that a significant second component in the X-ray spectra of BZ Cam and MV Lyr  is found that can be modeled
by a power law emission. As a result,
this detailed study on the three NL systems  concluded that the BLs in NLs can be optically thin hard X-ray emitting regions merged
with ADAF-like flows (advective hot flows) and/or constitute X-ray corona regions in the inner disk.
This interpretation is then consistent with non-detection of the soft X-ray  emission from a standard BL in a standard disk flow in a high state CV since
the inner region of the disk transits from a standard optically thick accretion flow into a non-standard flow resulting in advective hot flows (ADAF-like) in the X-ray emitting region.
Note here that winds are known to emit hard X-rays as a result of line-driven shocks. However, the detected temperatures and the luminosities are inconsistent with such an origin for
hard X-ray production as there is not enough mass loss to account for the X-ray luminosities (this is true for DNe, as well) \citep[see][for a discussion]{2014Balman-apj}.

This result predicts that the WDs should be {\it advectively heated} to higher temperatures as compared with WDs in binary systems
that do not have disks. \citet{2014Balman-apj} estimate that the  emission inefficiency
(in the BLs) by a factor $\sim$  0.01 can be accommodated with advective heating of WDs comparing WD temperatures in Polars (MCVs that are diskless) and NLs at similar orbital periods.
ADAF-like accretion flows can aid production of fast collimated outflows (3000-5300 km s$^{-1}$ as in BZ Cam and V592 Cas) because ADAFs
have positive Bernoulli parameter (the sum of the kinetic energy, potential energy and enthalpy).
Recently, these three NLs (BZ Cam, MV Lyr, V592 Cas) have been observed with \nustar\ in the energy band
3-78 keV which revealed a broad iron line complex around 6-7 keV including absorption and emission features, reflection lines in nonequilibrium ionization conditions (Balman et al. 2020, in preperation). 
Example \nustar\ spectra are given in Figure 3. 
The results are consistent with advective hot accretion flows (ADAF-like)  in the X-ray emitting region.
For a general discussion on the X-ray emission and the interpretation of the nature of BLs  and X-ray emitting regions  in the NL systems,
see \citet{2014Balman-apj}.

\begin{figure}\label{fig:3}
\includegraphics[width=7.0cm,height=5cm,angle=0]{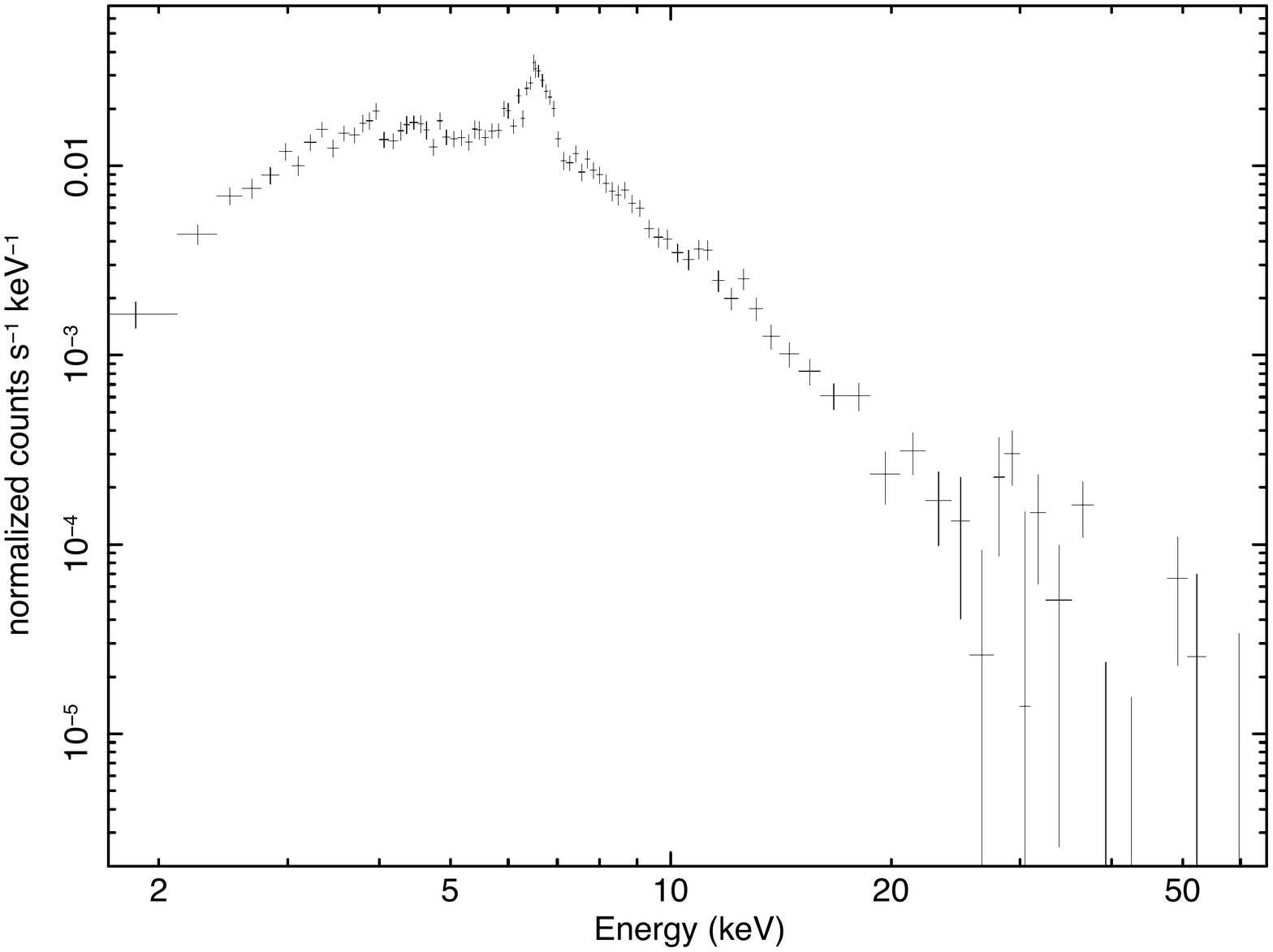}
\includegraphics[width=7.0cm,height=5cm,angle=0]{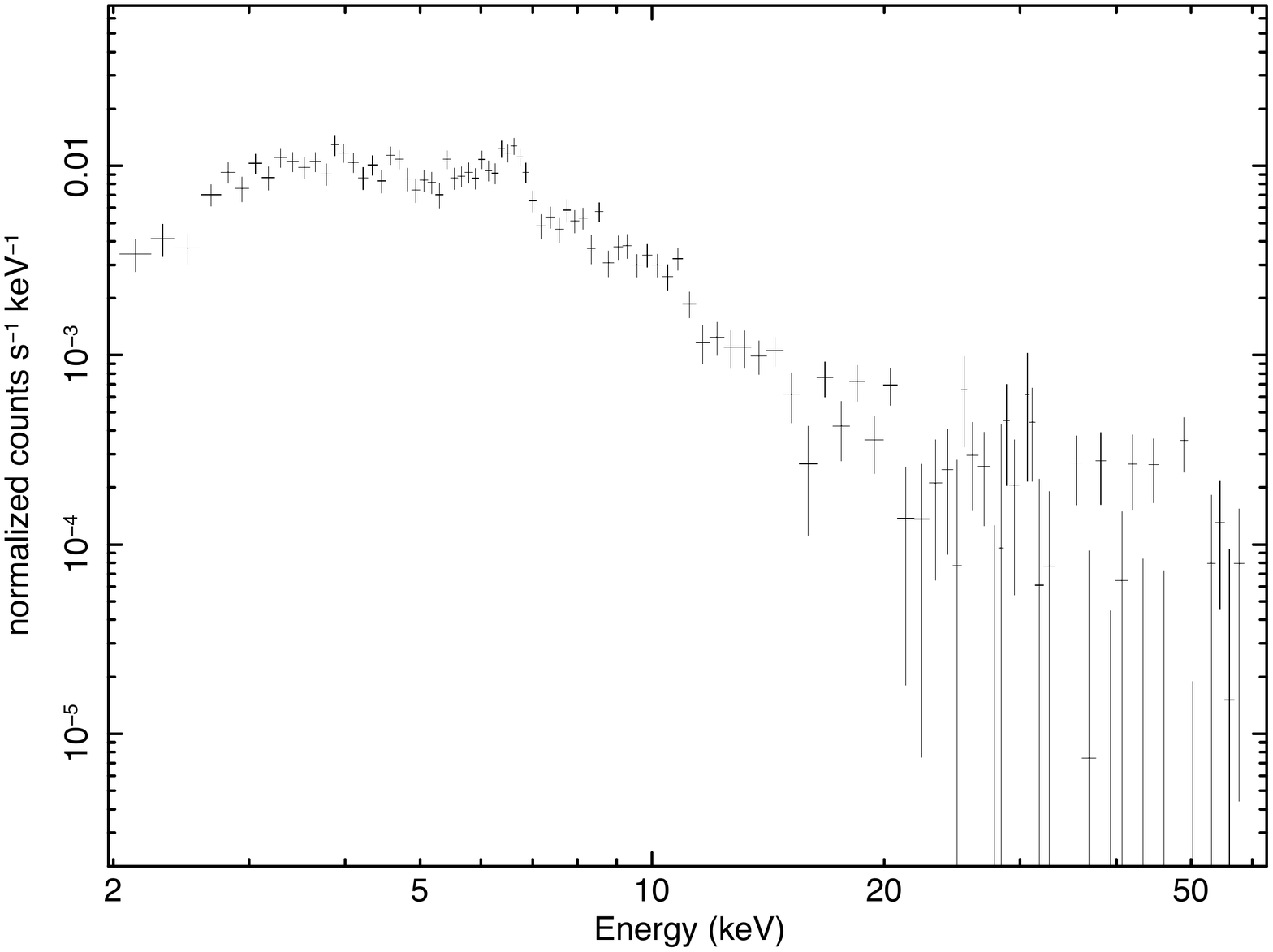}
\includegraphics[width=7.0cm,height=5cm,angle=0]{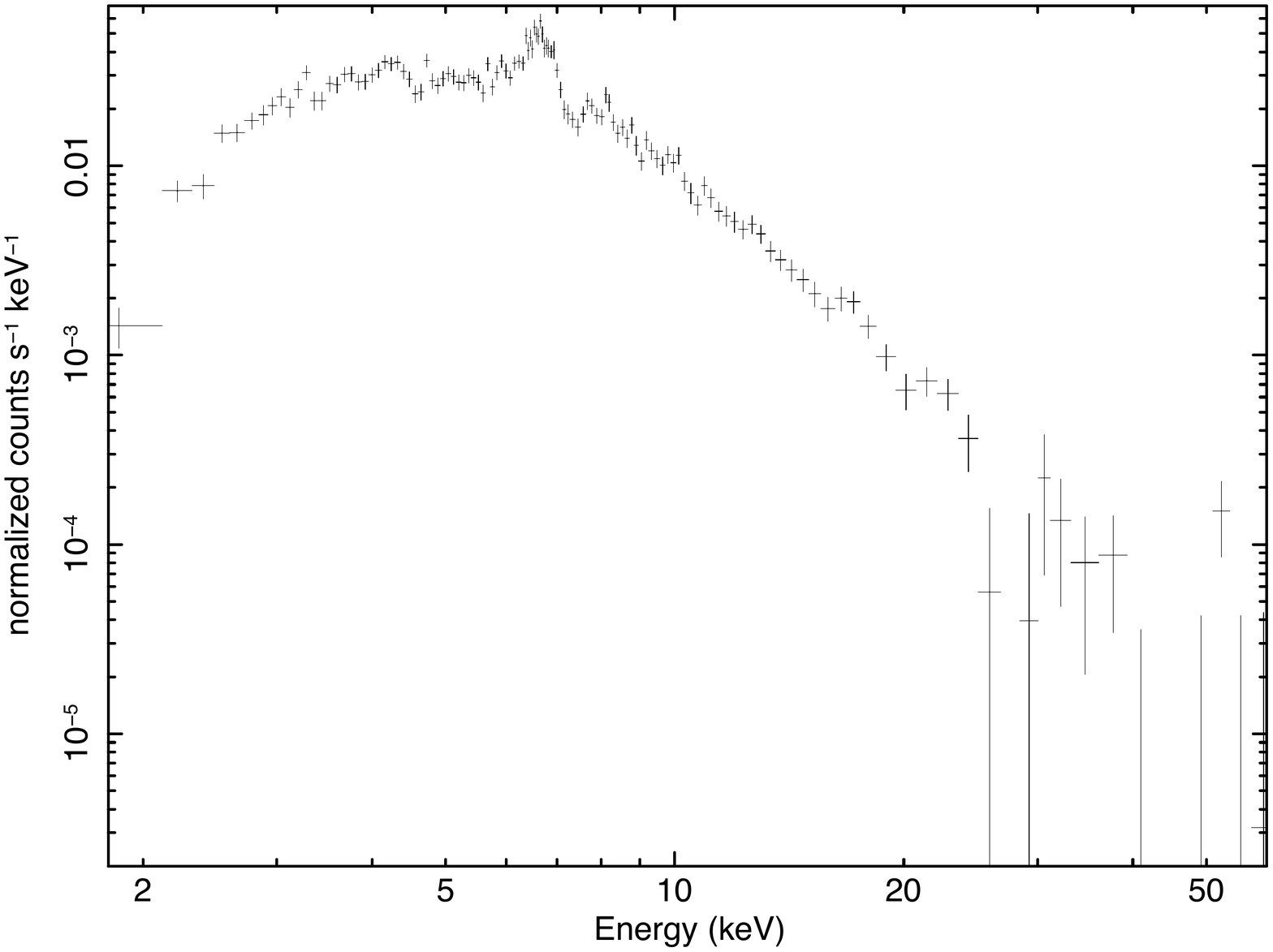}
\caption{Three nova-like spectra obtained by the \nustar\ satellite. From top to bottom : BZ Cam, V592 Cas, and MV Lyr. The crosses indicate
the spectral bins. A signal-to-noise ratio of 10 was used to bin the counts in the energy channels.}
\end{figure}

A study of the archival \chandra\  HETG data of CVs shows that the X-ray line emission is consistent with multi-temperature plasma in a
nonequilibrium state \citep{2014Schlegel} with n$_{e}$ between 10$^{12}$-10$^{16}$ cm$^{-3}$
which are expected  characteristics of advective hot flows. The UV (IUE and FUSE) data have also been found consistent with an inner truncation of the optically thick disk
using a large sample of NLs \citep{2017Godon}. In addition, UV spectroscopy of 33 NLs  have revealed
that the high accretion rate disks are departing from the standard disk models showing extended hot components \citep[][see also sec.(4)]{2007Puebla}.
A nonmagnetic NL, V3885 Sgr, has  been detected at 6 GHz with an optically thin synchrotron flux density of 0.16 mJy \citep{2011Kording}. \citet{2015Coppejans} have detected three other NLs  in the 
radio with a range of flux 0.03-0.24 mJy (using VLA) where the mechanism  is attributed to optically thick or thin synchrotron emission or electron cyclotron maser emission. 
Radio observations in compact accreting binaries (X-ray Binaries) are tracers of jets and certain flares
which are known to have strong connection to advective hot flows (ADAFs).

The best observational support for the existence of advective hot flows in DNe, during quiescence and outburst, comes from power spectral (timing) analysis and description of the broadband noise structure in optically thick 
disk flows and non-standard flows using the propagation of fluctuations model which will be discussed in the next Sec. (3).
Though many of the quiescent and outburst X-ray spectra of DNe seem consistent with isobaric cooling flow plasma emission in the X-rays, there are enough characteristics indicating that they constitute non-standard
BL regions that are merged with advective hot flows. Section (2.1) discusses some of these characteristics as sub-Keplerian velocities detected in Doppler broadening of lines and delays in the raise of UV and X-ray light curves in outburst modeled to show inner disk truncation.  Moreover, the X-ray luminosity of the emission  lines detected in DNe in quiescence and outburst are  low by about a factor between 10-100 compared with the continuum luminosity revealing that the plasma is not radiative (under-ionized)  \citep[e.g.,][]{2002Szkody,2014Schlegel} and when modeled by a collisional equilibrium plasma  model (e.g., isobaric cooling flow model) yields near solar or under-abundant elemental configuration due to lack of emission in the lines
\citep[e.g.,][]{2005Pandel,2011Balman}.
The persistence of a hard X-ray component that evolves during the outburst also strongly indicates the connection with advective hot flows since such flow type would not mediate a 
{\it typical} soft X-ray emitting BL that would be expected from standard accretion flows (i.e., but perhaps a thermal layer between the disk flow and the WD)  . 
Note the existence of a handful of soft X-ray emission from DNe during outburst in sec. (2.1) along with the hard X-ray component that is always there and evolves in the outburst.
A large sample ($\sim$ 722) of DN 
have been studied using the CRTS (Catalina Real Time Survey) yielding a median duty cycle of 5.8\% and recurrence time of 138 days \citep{2016Coppejans-st}. This average low duty cycle, and recurrence time is atypical for DIM expectations, but can be explained in the framework of  advective hot flow structure in the disk. 
SS Cyg is the brightest DN observed readily over the entire electromagnetic spectrum.  
Radio emission from SS Cyg during several  outbursts is interpreted as synchrotron emission originating from a transient jet and more recently other radio detections (15-80 ${\mu}$Jy 
in the range 8-12 GHz) of about 6 DN in outburst suggest existence of collimated jet-like outflows or flares as in XRBs 
\citep{2008Kording,2016Russell,2016Coppejans,2020Coppejans} known to have connection to advective hot flows.
In addition, the outburst spectra of WZ Sge obtained by the \chandra\ ACIS data show two component spectrum
of only hard X-ray emission, one of which may be fitted with a power law suggesting
thermal Comptonization of the optically thick disk photons and/or scattering from
an existing wind during the outburst (photon index varies from 0.8-2.0 and evolves in time). The thermal plasma component
evolves from 1.0 keV during the peak to about 30 keV after outburst. The spectral evolution and disappearance
of the power law component ($\simeq$10$^{30}$ erg s$^{-1}$) after outburst support the existence of nonstandard advective hot flows in the accretion disk within the X-ray emitting region \citep{2015Balman}.  

Advection-dominated accretion may be described in two different regimes.
The first is when the accreting material has a very low density
and a long cooling time (also referred as RIAF-radiatively inefficient accretion flow). This causes the accretion flow
temperatures to be virialized in the ADAF region at least in a nonequlibrium ionization condition.
Thus, RIAF ADAFs correspond to a condition where the gas is radiatively inefficient and the
accretion flow is underluminous \citep{2008Narayan,2008Lasota,2007Done}.
For CVs (and rest of AWBs, see sec.[2 and 6])
virial temperatures  in the disk are around 10-45 keV (assuming 0.4M$_{\odot}$ - 1.1M$_{\odot}$ WDs) which are typical values detected in quiescence, outburst and high accretion-rate states.
The accretion flow in this regime becomes
geometrically thick (extended), with high pressure support in the radial direction which causes
the angular velocity to stay at
sub-Keplerian values, and  the radial velocity of the gas becomes
relatively large with $\alpha$ (viscosity parameter) $\sim$0.1-0.3 (see also sec. [3]). Finally, the gas with the large velocity and scale height
will have low density, since the cooling time is long , and the medium will be optically thin. Note here that, the ADAF-like flows (advective hot flows)  in AWBs 
do not have the extreme  conditions as in the gravitational potential well of a black hole under severe general relativistic effects where the involved 
energies and flow characteristic would differ.
The second regime of ADAFs require that 
the particles in the gas can cool effectively,
but the scattering optical depth of the accreting material is large enough
that the radiation can not escape from  the system
\citep[see also "slim disk" model][]{1988Abramowicz}.
This regime requires high accretion rates of the order of 0.1$\dot{\rm M}_{\rm Edd}$.
The mass accretion rates derived from the optical and UV observations
for high state CVs are below this critical limit for a slim disk approach and at such high rates WDs are found as Super Soft X-ray sources 
burning the accreted hydrogen over their surface.  

\section{Aperiodic time variability of accretion flows and broadband noise in CVs}

Conventional flickering studies of CVs have been conducted using eclipse mapping techniques.
Some of these studies in quiescent DNe indicate that mass accretion rate
diminishes by a factor of 10-100 and sometimes by 1000 in the inner regions
of the accretion disks as revealed by the brightness temperature calculations
which do not find the expected R$^{-3/4}$ radial dependence of brightness temperature
expected from standard steady-state disks \citep[see][for a review]{2016Baptista,2019Balman}. 
Note that systems are more consistent with expectations during the DN outburst states.
On the other hand, this flattening in the brightness
temperature profiles may be lifted by introducing disk truncation in the quiescent state
\citep[e.g., r $\sim$ 0.15R$_{L1}$ $\sim$4$\times$10$^{9}$ cm; DW UMa, a nova-like:][]{2000Biro}.
A comprehensive UV modeling of accretion disks
at high accretion rates in 33 CVs including several nova-likes and old novae
\citep{2007Puebla} indicate an extra component from an extended optically
thin region, wind, or a corona/chromosphere evident from
the strong emission lines and the P Cygni profiles. This study calculates that the mass accretion
rate may be decreasing 1-3 orders of magnitude in the inner disk region.

Another method used in flickering studies  of accreting objects  
is the aperiodic variability of brightness (resulting in broadband noise of power spectra) of 
sources in the X-rays which may be used in structural diagnosis in the accretion disks.
Generally, the long time-scale variability is created in the outer parts of the accretion disk
\citep{1971warner}, and the relatively fast time variability
(at $f>$few mHz) originates in the inner parts of
the accretion flow  \citep{2004Baptista,2000Bruch,2015Bruch,2016Baptista}.
Properties of this broadband noise is similar to that of the X-ray binaries with neutron stars and black holes.
The widely accepted model for this aperiodic flicker noise is a model of propagating fluctuations
\citep{1997Lyubarskii,2009Revnivtsev,2010Revnivtsev,2011Uttley,2013Ingram,2016Ingram}.
The modulations of light are created by variations in
the instantaneous value of the mass accretion rate in the region of the energy release.
These variations in the mass accretion rate, in turn, are inserted into the flow at all Keplerian radii throughout the disk
 due to the stochastic nature of
its viscosity and then transferred toward the compact object. These variations are characterized with a multiplicative 
time series on dynamical timescales.
This model predicts that the truncated optically thick accretion disk should lose its variability characteristic at
high Fourier frequencies. Thus, the nature and the expected model of the broadband noise should change.

The truncation of the optically thick accretion disk in DNe in
quiescence was already invoked as a possible explanation for the time lags between the optical and UV
fluxes in the rise phase  of the outbursts \citep{1994Meyer,1999Stehle},
and for some implications of the DIM \citep[see][]{2004Lasota}
or due to the unusual shape of
the optical spectra or light curves of nonmagnetic CVs \citep{2005Linnell,2011Kuulkers}.
SS Cyg is the brightest DN observed readily over the entire electromagnetic spectrum.  

A recent paper by \citet{2019Balman} reviews the properties of the broadband noise in CVs, mainly DNe, discussing the spectral timing 
characteristics as to state changes together with anticorrelations of break frequencies with thermal temperature in the X-ray emitting zone in the context of advective hot flows. 
\citet{2012Balman} have used the broad-band noise characteristics
of selected DN in quiescence (only one in outburst: SS Cyg)
and studied the inner disk structure and disk truncation via propagating fluctuations model.
The power spectral densities (PDS) were calculated in terms of the fractional
rms amplitude squared following from \citep{1991Miyamoto} and expressed in units of $(\rm{rms}/\rm{mean})^2$/Hz.
This was multiplied with the frequencies to yield
${\nu} P_{\nu}$ versus ${\nu}$, integrated power. The broad-band noise structure of the Keplerian disks often
show $\propto$ $f^{-1\dots-1.3}$ dependence on frequency
\citep{2001Churazov,2005Gilfanov}, and this
noise  shows a break if the optically thick disk truncates as the Keplerian motion
subsides/changes typical characteristics. \citet{2012Balman} show that for five
DN systems, SS Cyg, VW Hyi, RU Peg, WW Cet and T leo, the UV and X-ray power
spectra show breaks in the variability with break frequencies in a range 1-6 mHz (see Figure 4 and Table 1),
indicating inner optically thick disk truncation in these
systems. The truncation radii for DN are calculated in a range $\sim$(3-10)$\times$10$^{9}$ cm
including errors \citep[see Table 2 in][]{2012Balman}. \citet{2014Balman,2015Balman,2019Balman} presents
PDS analysis of four more DNe with relevant break frequencies (see Figure 4 and Table 1).

The same authors used the archival $RXTE$ data of SS Cyg in quiescence and
outburst and show that the disk moves towards the WD during the optical peak to
$\sim$ 1$\times$10$^{9}$ cm ($\sim$50 mHz)  and recedes
as the outburst declines to a quiescent location at 5-6$\times$10$^{9}$ cm ($\sim$5 mHz). This is shown
for a CV, observationally, for the first time in the X-rays (see Figure 5 top left panel).
A similar study in the optical by  \citet{2012Revnivtsev} reveals
very similar results to X-rays and no break during the optical peak out to a frequency of  100 mHz.

\begin{figure*}\label{fig:4}
\centerline{
\includegraphics[width=8cm,height=5cm,angle=0]{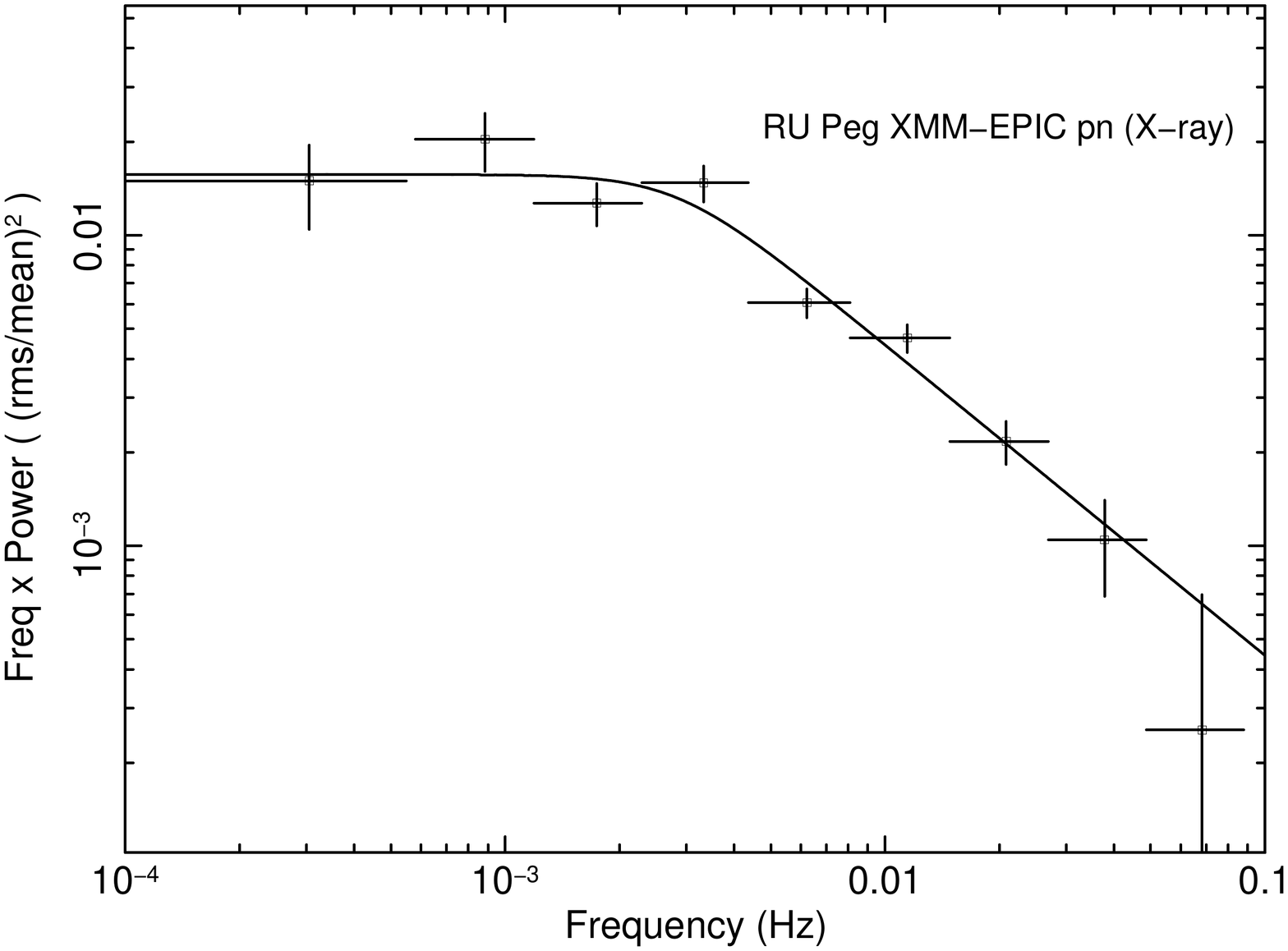}
\includegraphics[width=8cm,height=5cm,angle=0]{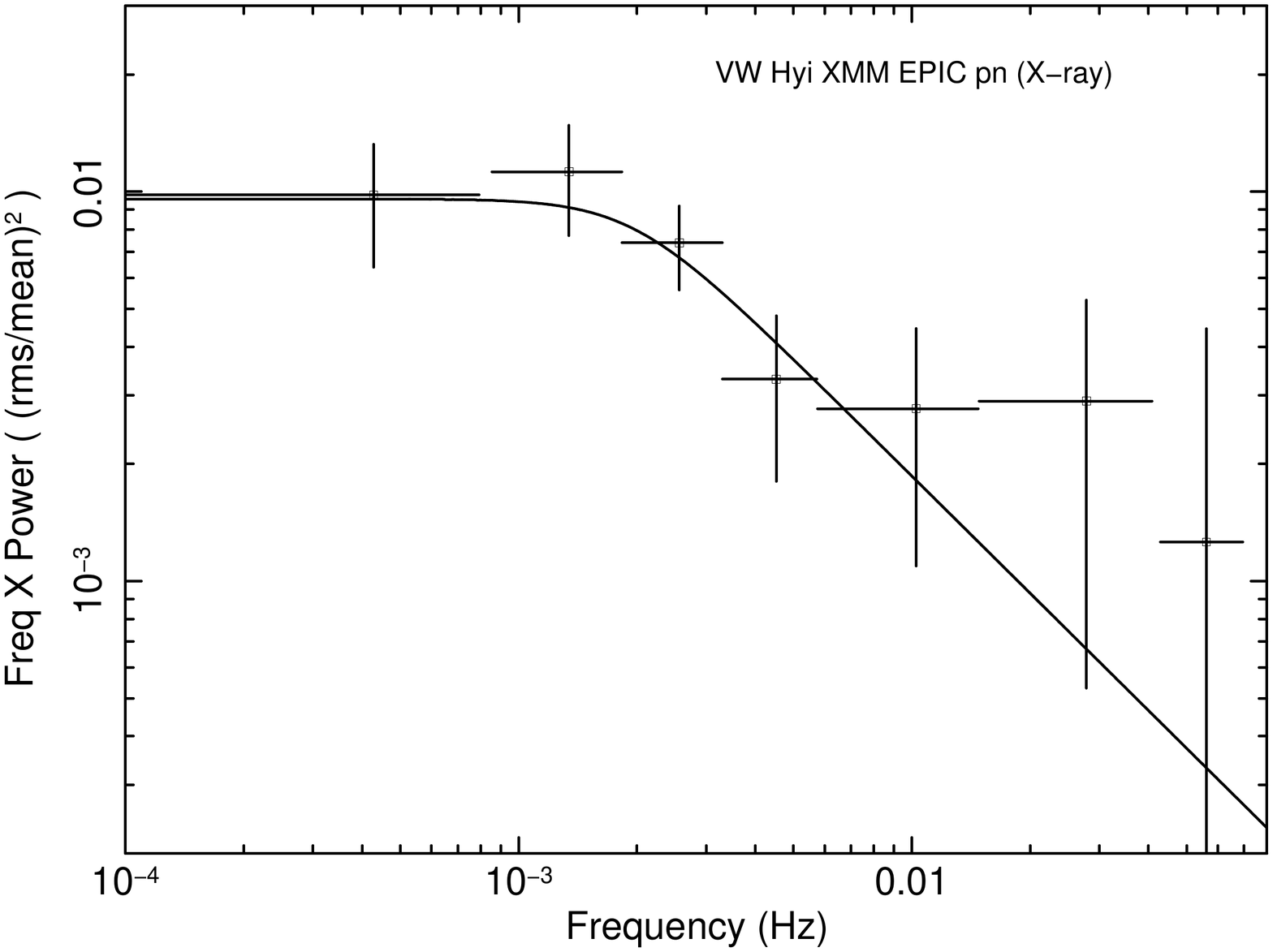}}
\centerline{
\includegraphics[width=8cm,height=5cm,angle=0]{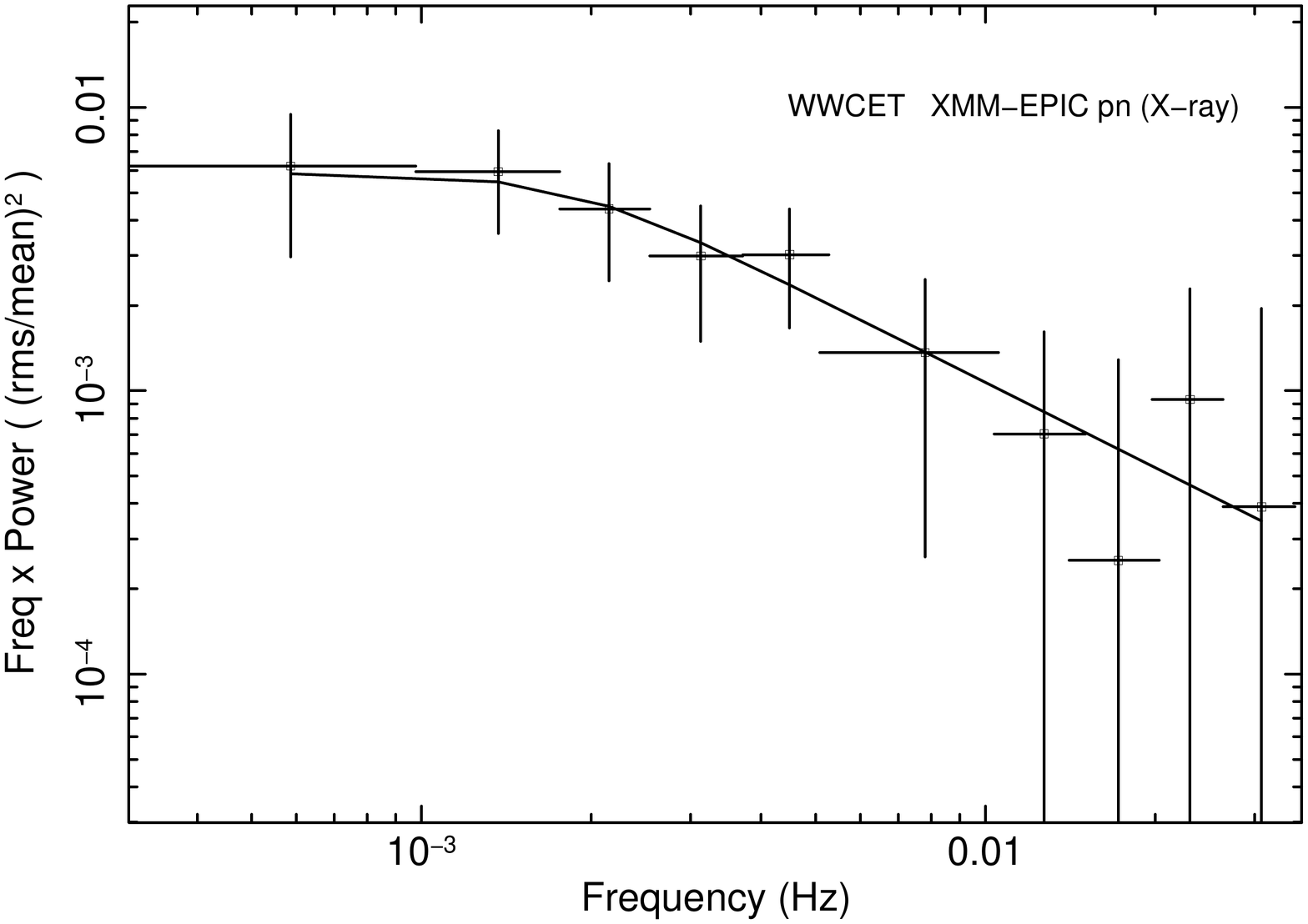}
\includegraphics[width=8cm,height=5cm,angle=0]{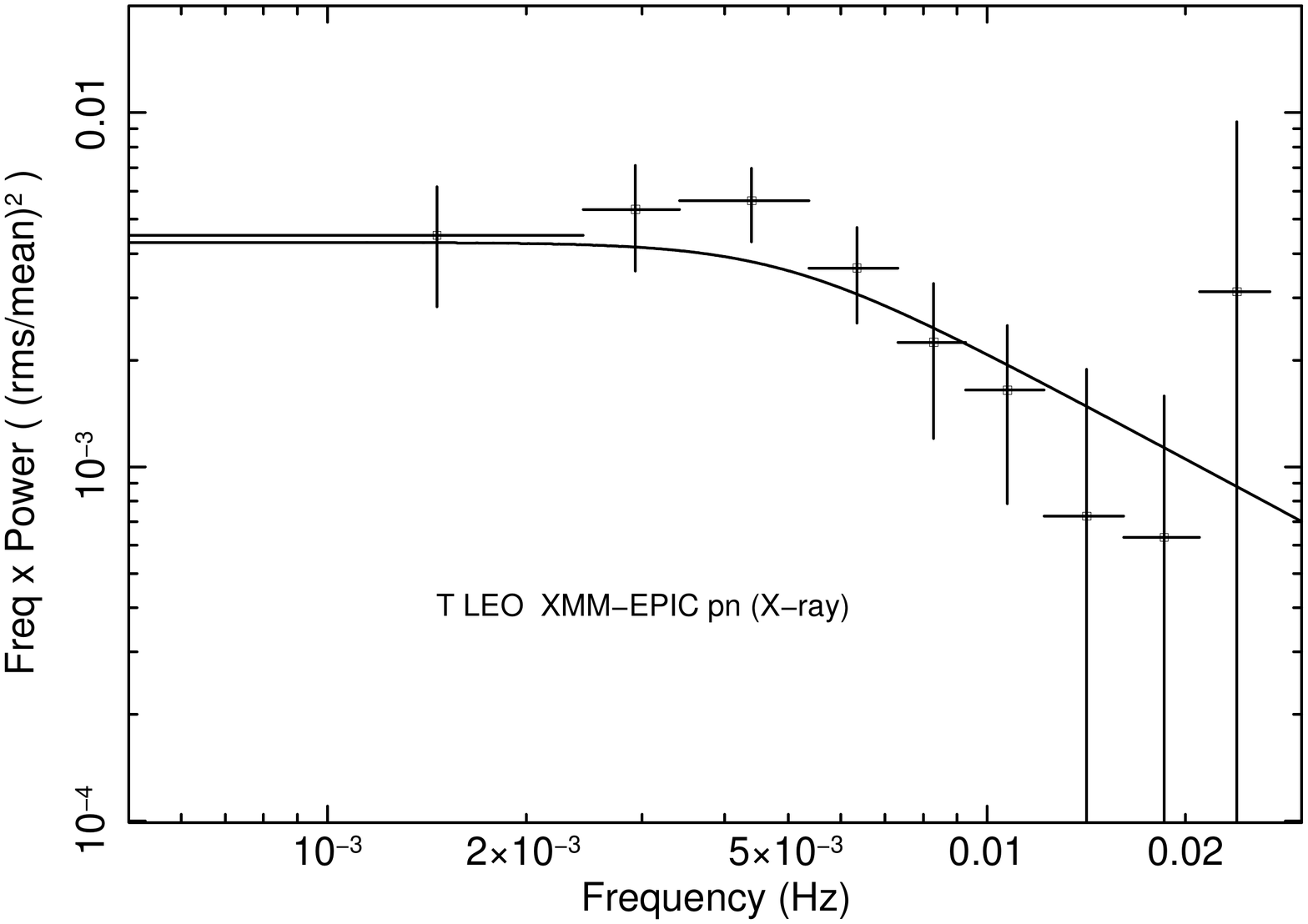}}
\centerline{
\includegraphics[width=8cm,height=5cm,angle=0]{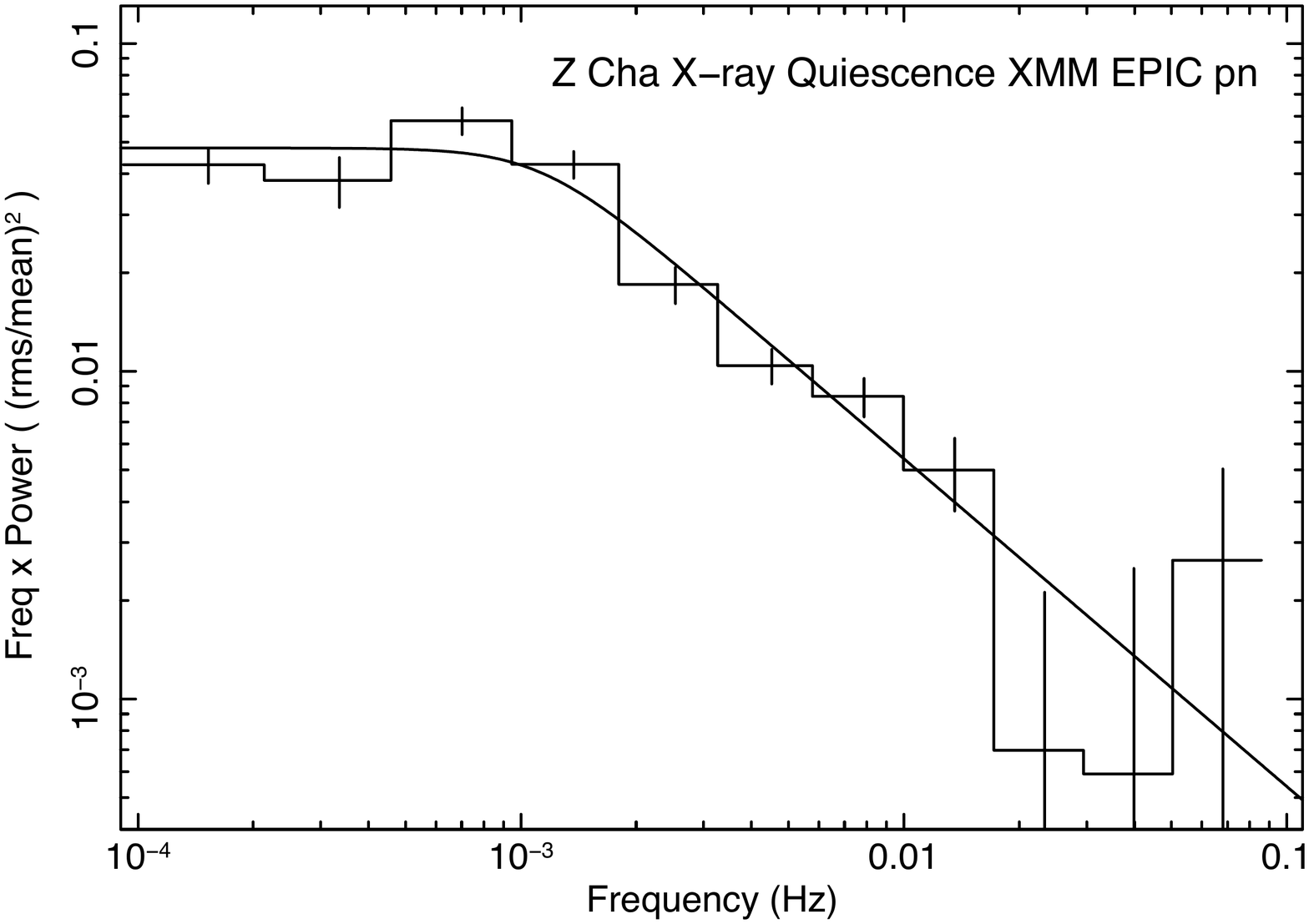}
\includegraphics[width=8cm,height=5cm,angle=0]{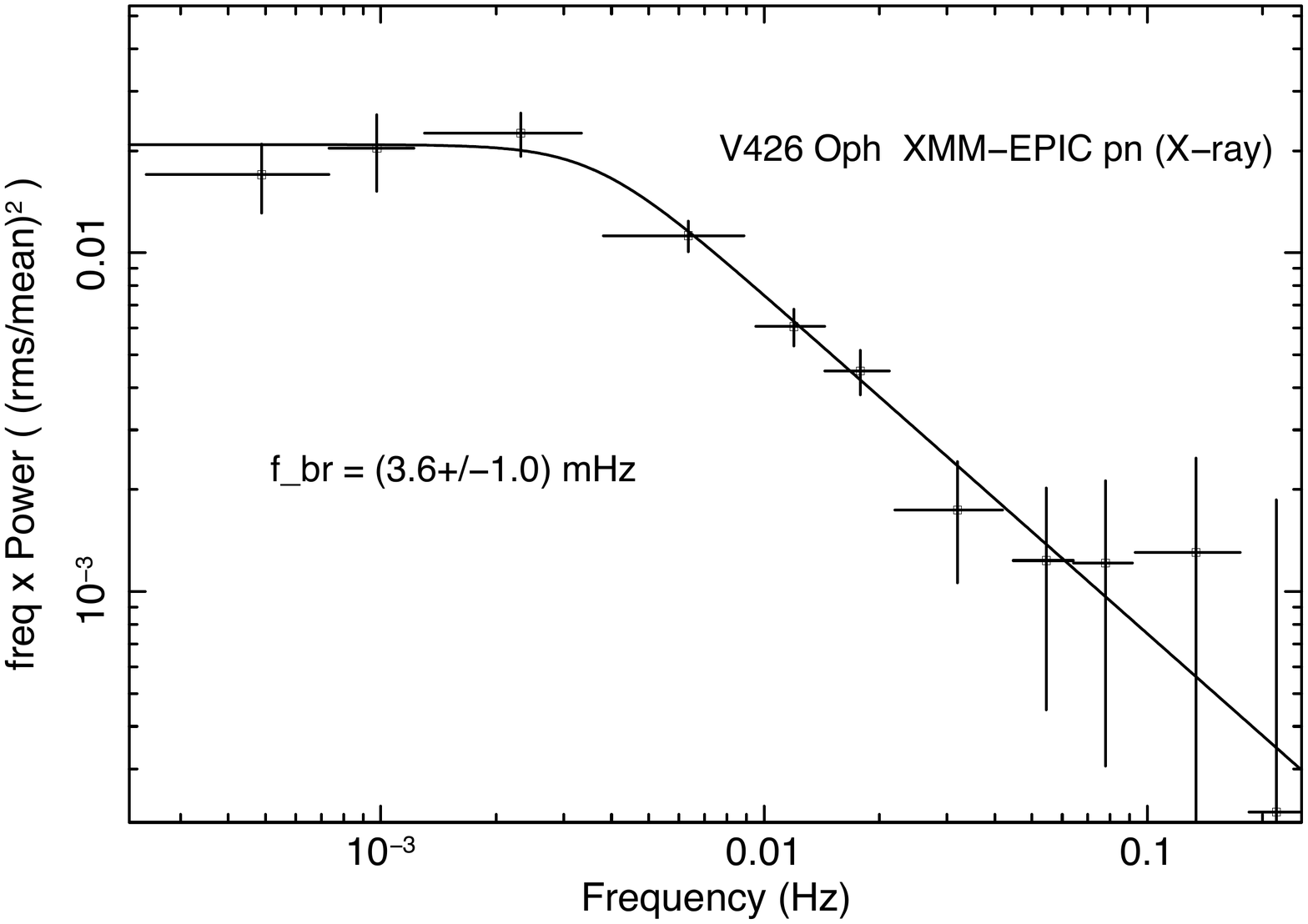}}
\centerline{
\includegraphics[width=8cm,height=5cm,angle=0]{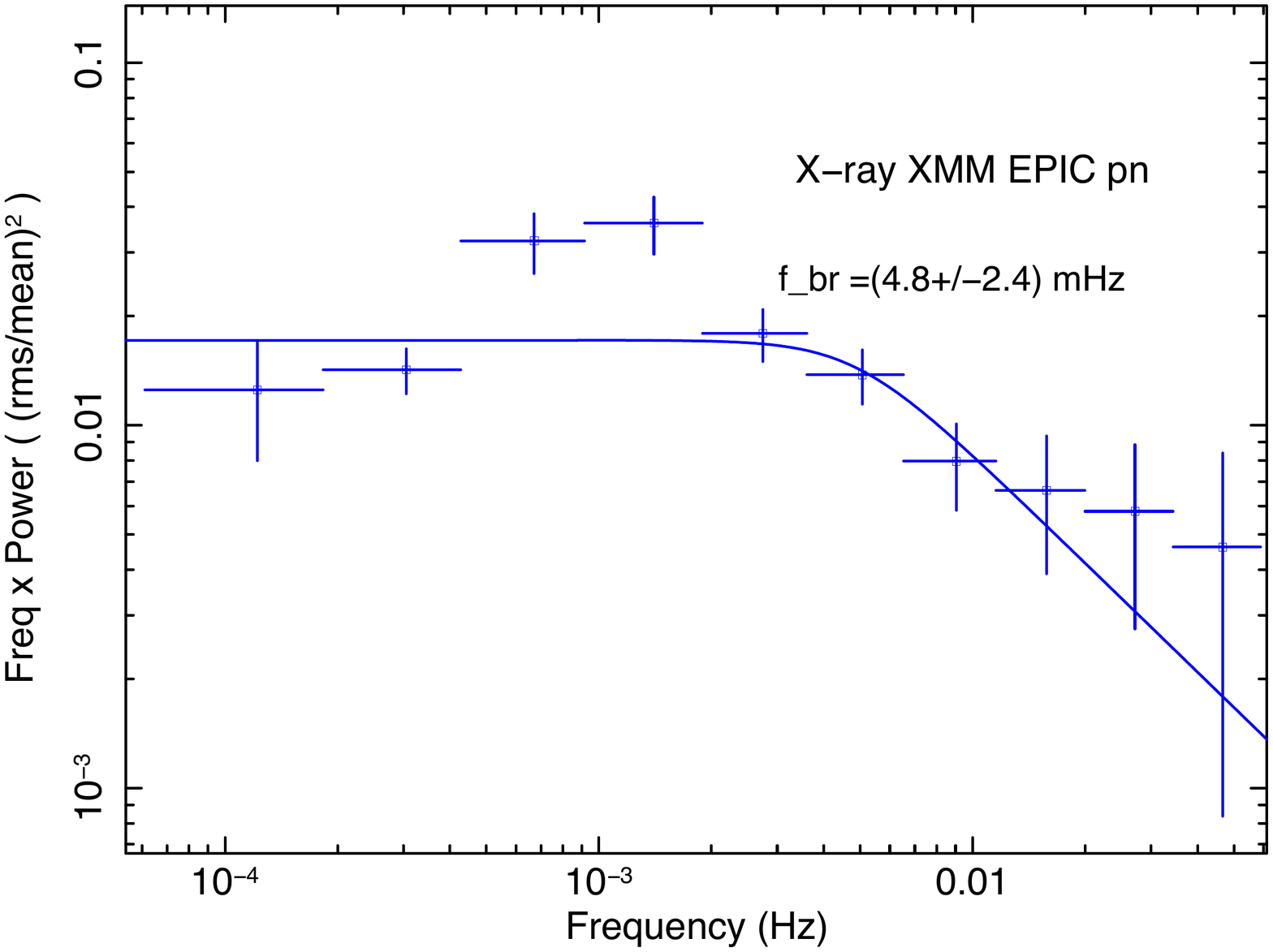}
\includegraphics[width=8cm,height=5cm,angle=0]{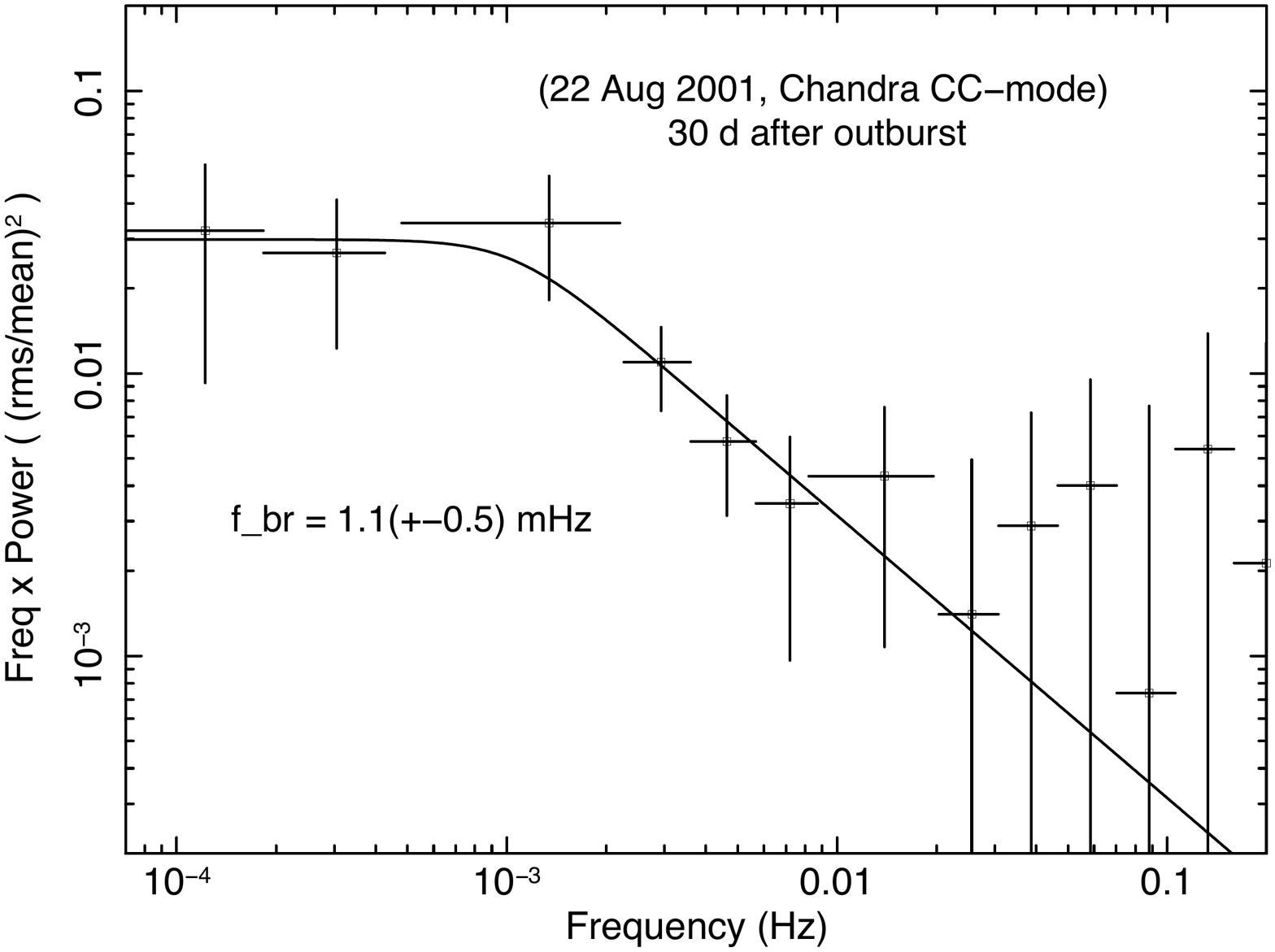}}
\caption{ PDS of some nonmagnetic CVs, DNe,  obtained during quiescence using \xmm\ data. Top left is RU Peg and top right is VW Hyi.
The second panel is WW Cet on the left  and T Leo on the right. Third panel has Z Cha on the left and V426 Oph on the right.  Bottom panel
displays HT Cas (on the left) and WZ Sge (on the right). 
The solid lines show the fit with the propagating fluctuations model for the PDS. Note the QPO like low significance peaks around the breaks in the PDS.
See \citet{2012Balman,2015Balman,2019Balman} for details of the analysis and the figures.}
\end{figure*}

Two other DN were analyzed in the same manner as SS Cyg in outburst and quiescence (see Balman 2015, 2019).  
SU UMa is  considered to be a  prototype of its subclass and 
WZ Sge is a short period SU UMa type dwarf nova with a long inter-outburst interval of 20-30 yrs
as opposed to the recurrence time of 12-19 d for SU UMa \citep{2010Collins}. 
Note that these two DN show different characteristics in their optical outburst light curves. SU UMa indicates an anticorrelation between X-rays and optical
light curves as one peaks the other is suppressed. However, WZ Sge shows no clear anticorrelation between X-ray and optical light curves, but the X-ray luminosity
increases towards quiescence.
There seems to be no break in the X-ray PDS out to a frequency of  100 mHz for SU UMa and  200 mHz for WZ Sge after which the noise in the PDS 
disappears during the optical peak of the
outburst.  Table 1 shows the relevant break frequencies where the approximate optically thick disk truncation occurs in the quiescent and outburst phases.
Note that the transitions occur approximately at 3.8$\times$10$^{9}$ cm for SU UMa and 1.3$\times$10$^{10}$ cm for WZ Sge during quiescence. 
Figure 5 shows the X-ray PDS of SU UMa in quiescence and outburst obtained using $RXTE$ data from \citet{2019Balman} where the PDSs of
WZ Sge is also presented.
Therefore, though the optical light curves of these sources indicate different characteristics and classifications for SS Cyg, SU UMa and WZ Sge,
the inner disk moves in all the way/close to the WD during the optical peak
and moves out in decline to quiescent location further out as detected using the X-ray PDSs. However, I note that there may be 
variations in the break frequency during quiescence as noticed in further PDS analysis of WZ Sge and other
DNe.

\citet{2010Revnivtsev} study the power spectra of the variability of seven
IPs  containing magnetized asynchronous accreting WDs
in the optical and three of them (EX Hya, V1223 Sgr, TV Col), both in the X-rays
and the optical \citep[see also,][]{2011Revnivtsev}. Their time variability and
broadband noise can be explained by
the propagating fluctuations model in a truncated optically thick accretion disk in a similar fashion to DNe
except that the truncation in IPs is caused by accreting material being channeled to the magnetic poles
of the WD and in DNe the physical conditions in the flow changes after the break and the accretion  reaches the WD at the end.
The question of the inner disk radius in IPs was also investigated for GK Per by \citet{2004Hellier}  and recently using
the similar formalism as Revnivtsev et al. by \citet{2019Suleimanov} to reveal differences for quiescence and outburst together with
relevance for mass determination.

Accretion-powered X-ray pulsars and asynchronous magnetic white dwarfs
(intermediate polars) have magnetic fields strong enough to disrupt
the inner parts of the accretion disks where the fastest variability
timescales associated with the innermost regions of the disk should be
absent or reduced in their power spectra. The authors' work show that
the PDS have breaks at Fourier frequencies associated with the
Keplerian frequency of the disk at the white dwarf magnetospheric boundaries as illustrated in Figure 6.
The values of the break frequencies for V1223 Sgr in the optical and
in X-rays are: $f_{\rm break, opt}=(21\pm5)$ mHz,
$f_{\rm break, \mbox{X-ray}}=(33.6\pm3)$ mHz. The break frequencies are compatible at the 2$\sigma$ level
for the optical and the X-rays and yield disk truncation at around r$_{\rm m}$$\sim$1.6$\times$10$^{9}$\ cm. This truncation radius is around
r$_{\rm m}$$\sim$2$\times$10$^{9}$\ cm for EX Hya. \citet{2011Revnivtsev} show that in three cases the PDS of flux variability
in the X-ray and optical bands are similar also for EX Hya and 
and the majority of X-ray and optical fluxes are correlated with time
lag $<$ 1 s. Thus, the variable component of the optical emission from the
accretion disks in these binary systems originate from a component that is due to the
reprocessing of the X-ray luminosity in these systems.
In the EX Hya data, they  detect
that optical emission leads the X-ray emission
by about 7 s. The authors interpret this in the framework of the model of propagating
fluctuations consistent with the travel time of matter from the
truncated accretion disk to the white dwarf surface with a magnetically channeled flow. \citet{2019Suleimanov} finds that the changes of the X-ray temperatures of  the magnetic dwarf novae 
(EX Hya and GK Per) from quiescence into outburst, are a result of  the change in  the break frequencies and the changing inner disk radius . This indicates that magnetospheric radius  
changes with accretion rate (the optically thick disk moves in during outburst and pulls out
towards the quiescence, changing the magnetospheric size). Note that, as reviewed by \citet{2019Balman}, such changes of the disk moving in and out can occur during quiescence, as well.

Some optical band studies using $Kepler$ data \citep{2012Scaringi}
for the nova-like CV MV Lyr
reveals that the source has log-normal flux distribution in the
rms-flux relation and the origin of variablity is the multiplicative processes
travelling from the outer to inner disk \citep[as opposed to simple additive processes, e.g.,][]{2014Dobrotka}\ 
proposed by the propogating fluctuations model mentioned at the beginning of this section.
The long term $Kepler$ data analysis of MV Lyr shows that
all PSDs indicate single or several
quasi-periodic oscillations (QPOs) along with
a frequency break around 1-2 mHz. This frequency break, at about 1-2 mHz, may be similar in origin to the
the breaks observed in DNe in quiescence. However, the analysis of X-ray PDS using the \nustar\ data shown in the previous
section reveals that there is a constant broadband noise out to 10 mHz at the level of 0.001-0.002 (rms/mean)$^2$ after which the noise disappears.
\citet{2013Scaringi} calculates simultaneous
optical lightcurves in differents bands, and finds soft lags of 3-10 sec where the blue
photons are observed before the red ones, with larger lags at
low frequencies. This may be related to the reprocessing of harder radiation in the
UV and X-ray regimes by the outer cooler disk on possibly thermal timescale. Similar soft lags have been determined for
SS Cyg with similar origins \citep{2018Aranzana}.

\begin{figure*}\label{fig:5}
\centerline{
\includegraphics[width=7cm,height=5cm,angle=0]{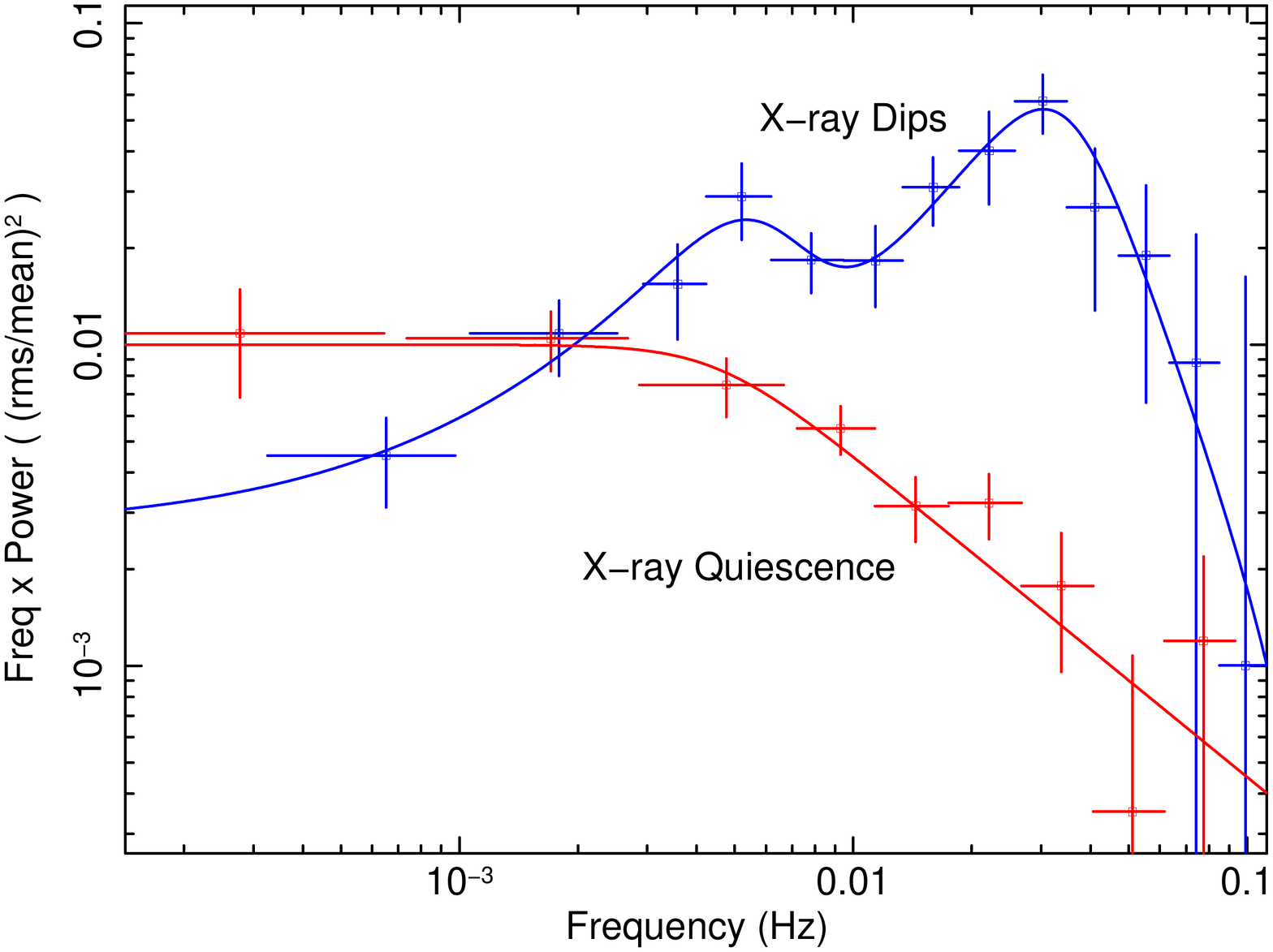}
\includegraphics[width=7cm,height=5cm,angle=0]{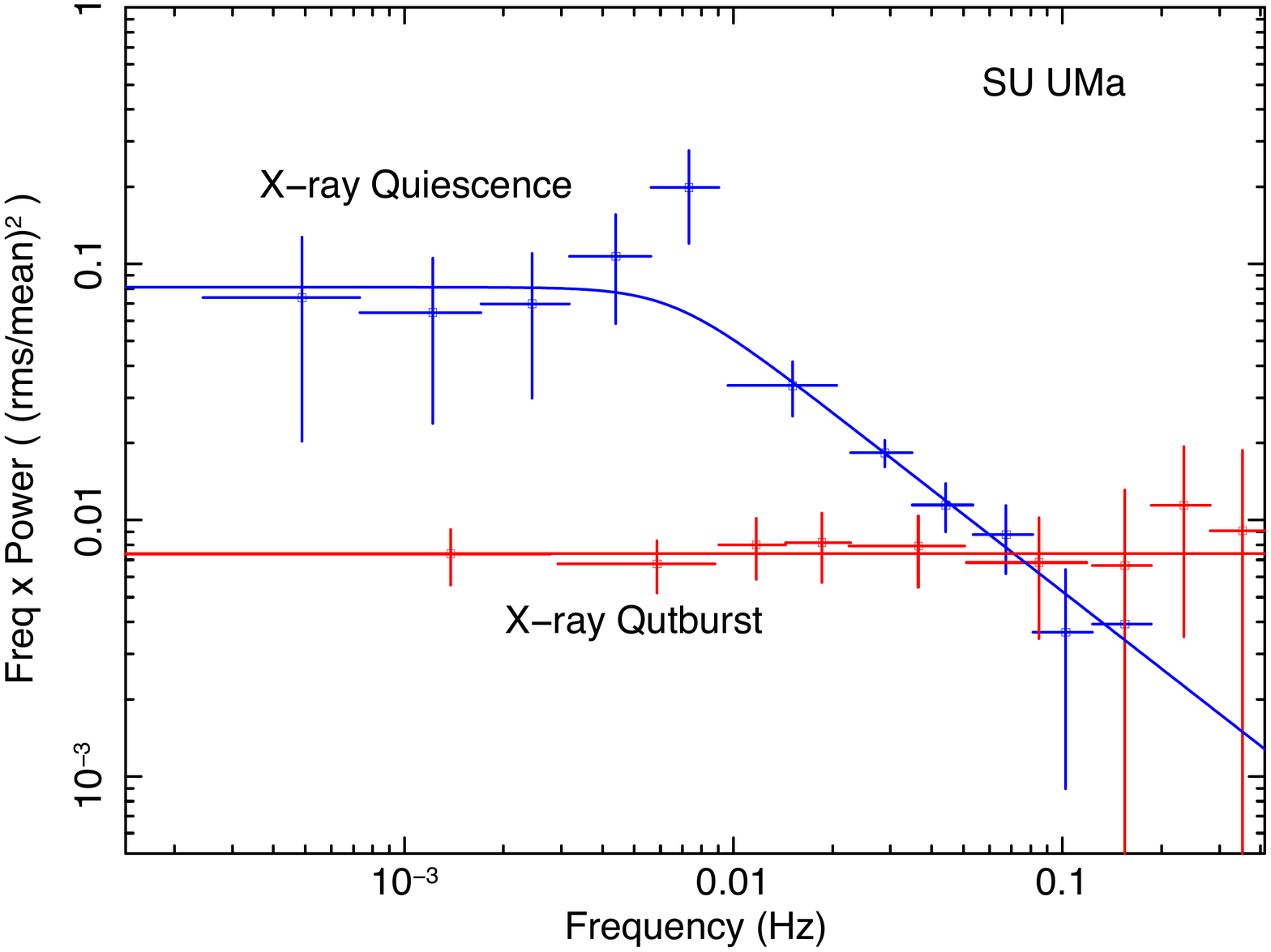}}
\caption{PDS of SS Cyg  on the left-hand (\rxte\ data) and SU UMa on the right-hand side (\rxte\ data).
See \citet{2012Balman,2015Balman,2019Balman} for details of the PDS.
The X-ray PDS are obtained from data in the outburst stage (optical peak and X-ray suppression) and quiescence .
The solid lines show the fit with the propagating fluctuations model.  For the PDS of the X-ray dips/suppression of SS Cyg
two Lorentzians along with the propagating fluctuations  model were used to achieve the best
fitting results. }
\end{figure*}

\begin{figure*}\label{fig:6}
\centerline{
\includegraphics[width=8cm,height=5cm,angle=0]{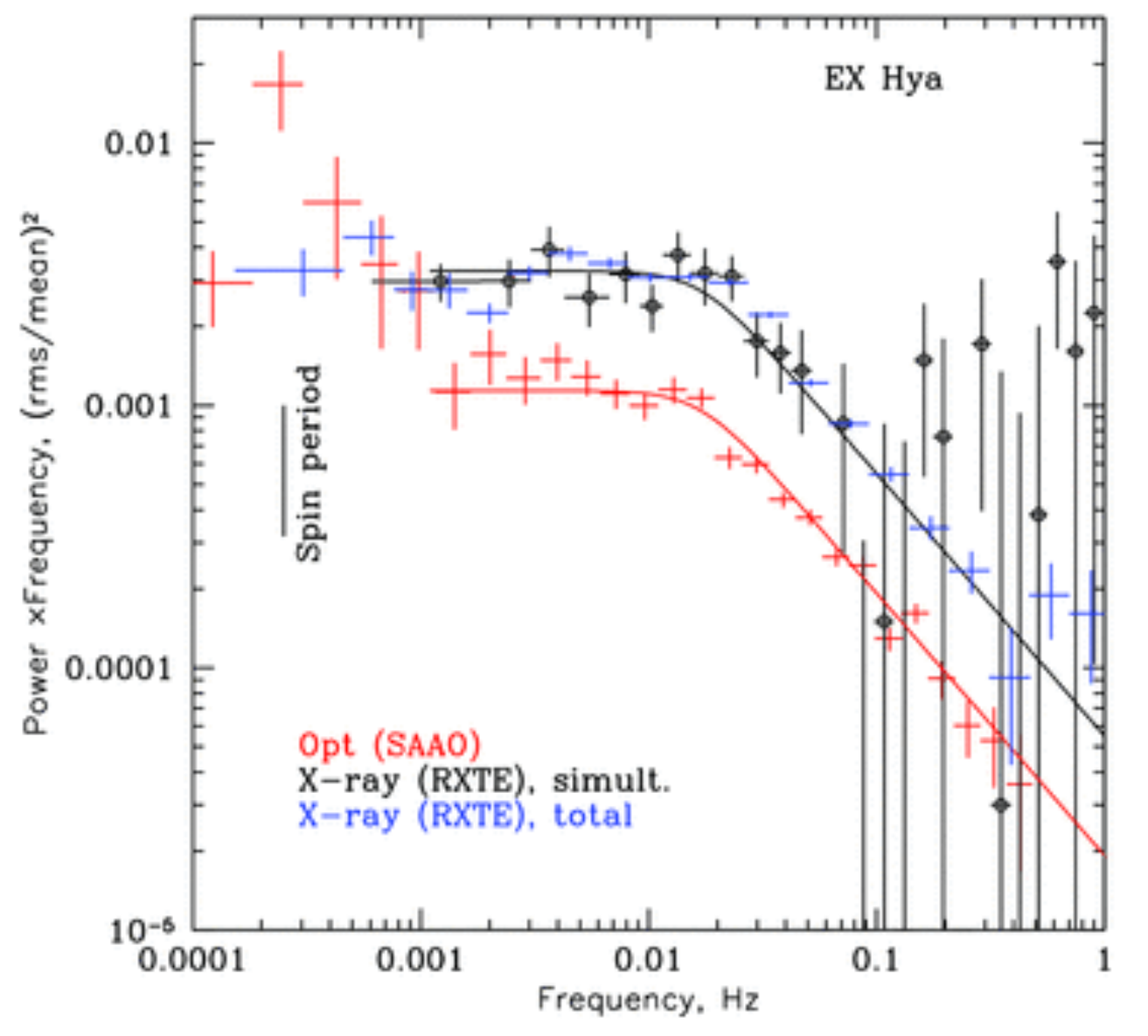}
\includegraphics[width=8cm,height=5cm,angle=0]{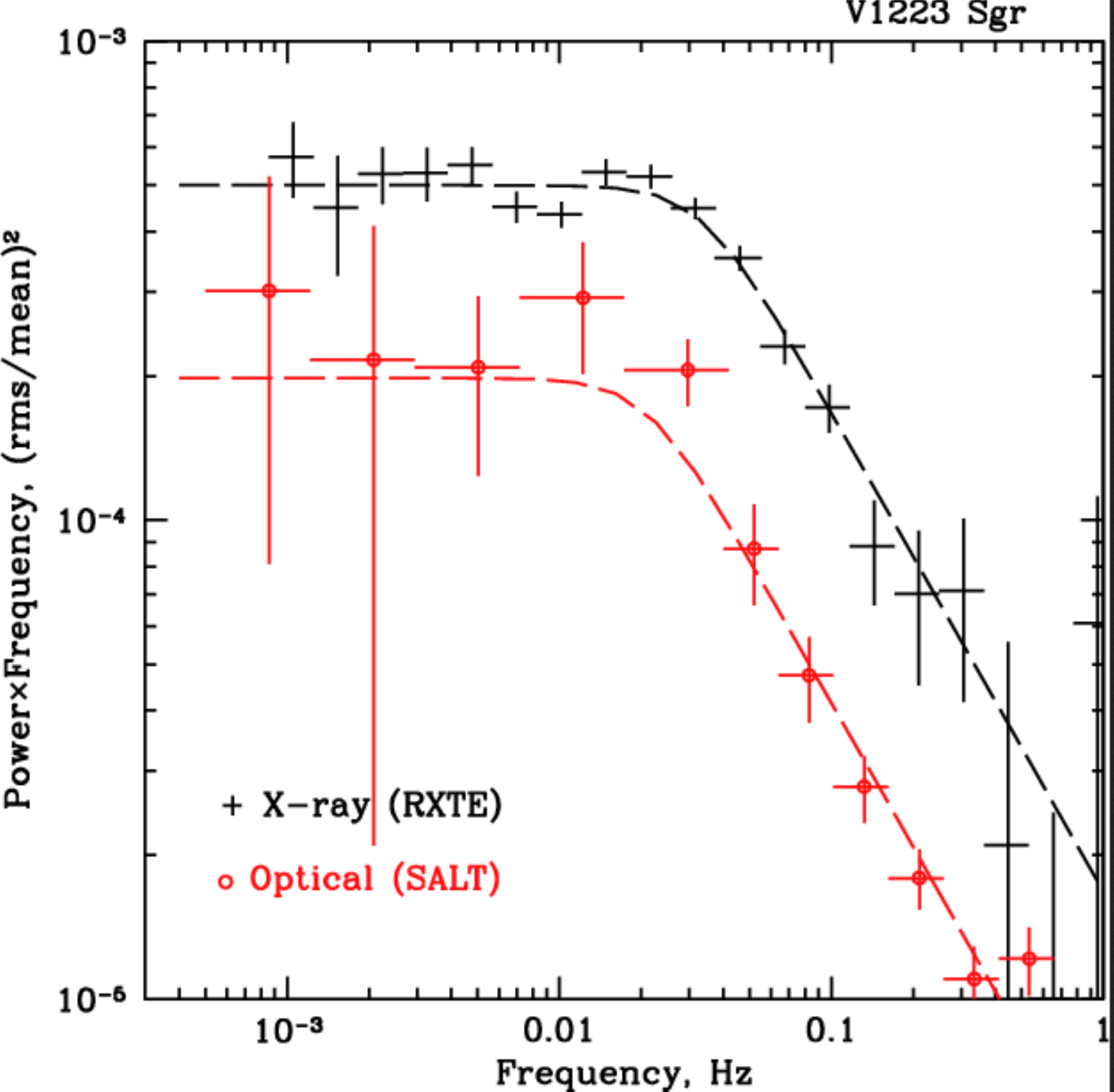}}
\caption{PDS of EX Hya on the left hand side and V1223 Sgr on the right hand side. Both X-ray and optical PDS
have been presented. Data obtained simultaneously or at different times have been labeled accordingly on the figure. 
The solid lines show the fit with the propagating fluctuations model. See \citet{2011Revnivtsev} for details of the analysis and PDS produced.}
\end{figure*}

If the model of the propagating fluctuations correctly describes the time variability of
the X-ray and optical/UV flux of DNe, then we should expect some particular way to correlate the
brightness of systems at these energies. But the optical/UV light
variations, generated as an energy release of variable mass accretion rate at the inner edge
of the optically thick accretion disk flow, should lead the X-ray emission with the time lag equal to the
time needed for the matter to travel from the inner edge of the optically thick flow to the
central regions of the accretion flow in the vicinity of the WD, where the bulk of the
X-ray emission is generated. \citet{2012Balman} make cross-correlation analysis of
the X-ray and UV data of four DNs by calculating CCFs (see also Figure 7). 
The CCFs for all the dwarf novae show  clear asymmetry indicating that some part of
the UV flux is leading the X-ray flux. In addition, a strong peak is detected
near zero time lag for RU Peg, WW Cet and T Leo (also Z Cha, Balman 2020 in preparation), suggesting a significant zero-lag correlation between the X-rays
and the UV light curves. The positive time lag, leading to an  asymmetric profile for the five novae above
and the shifted profile for SS Cyg show that the X-ray variations are delayed relative to those in the UV as expected from the propagating fluctuations model
and the inner advective hot flow characteristics. The zero time lag for all the systems shows that there is significant reprocessing going on in DNe which is also
consistent with the soft-lags detected for some systems in the optical wavelengths. 

Overall, peaks near zero time lag in the cross-corelation studies (for six systems) showing light travel
effects indicate that part of the UV emission arises from reprocessing of the X-ray emission originating from the inner ADAF-like flow.
X-ray time lags/delays with respect to the UV wavelengths (7 DNe) detected in quiescence at 96-209 s are consistent with
with matter propagation timescales onto the WD in a truncated optically thick
nonmagnetic CV disk revealing the existence of advective hot flow regions with a different characteristics  (emissivity, viscosity, radial and azimuthal velocity, etc.).
An $\alpha$ (viscosity parameter) of 0.1-0.3 can be estimated for the inner regions of the DNe accretion disks in
quiescence using comparative time lags between the X-ray and the UV or optical light curves
of DNe and MCVs (IPs). 10 DN systems are successfully modeled with a power law index  $\alpha$=1 and  after the break in the frequency  $\alpha$=2. 
In the outburst, the systems show the $\alpha$=1 component out to around 50-200 mHz after
which there is no noise detected. In this stage, the PDS analyses indicate that the rms (\%)
variability diminishes as expected since the optically thick disk is supported by radiation 
in the three cases studied SS Cyg, SU UMa, and WZ Sge (only SS Cyg shows soft X-ray emission). 

In general, the disk truncation in the MCVs that are IPs, is due to the magnetic channeling of the
accretion flow as the ram pressure and magnetic pressure becomes compatible. The break frequency and truncation  radii are, generally, smaller
than in the nonmagnetic DNe systems in quiescence ((0.9-2.0)$\times$10$^9$\ cm; about 10 mHz). 
The values of the break frequencies in the PDS of IPs
can be used to make estimates of the inner radii
of the truncated accretion disks and the white dwarf magnetic fields.
The results for IPs are analogous to NS XRBs
with magnetic fields that yield disk truncation. The break frequency resembles that of  the spin frequency in
corotating pulsars which strongly suggests that the typical variability timescale in accretion disks
is close to the Keplerian frequency.

\begin{figure*}\label{fig:7}
\centerline{
\includegraphics[width=8.1cm,height=4.3cm,angle=0]{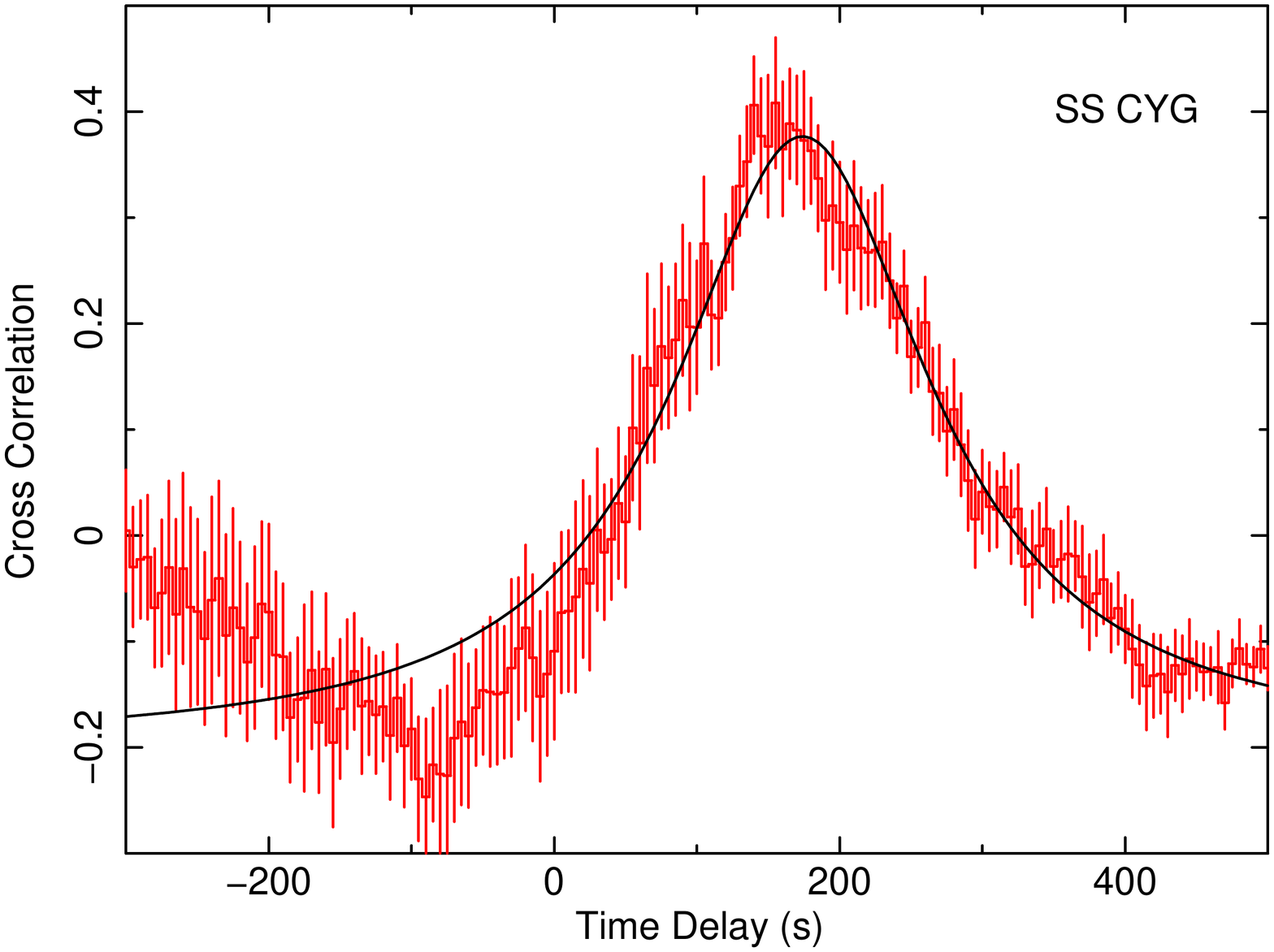}
\includegraphics[width=8.5cm,height=4.3cm,angle=0]{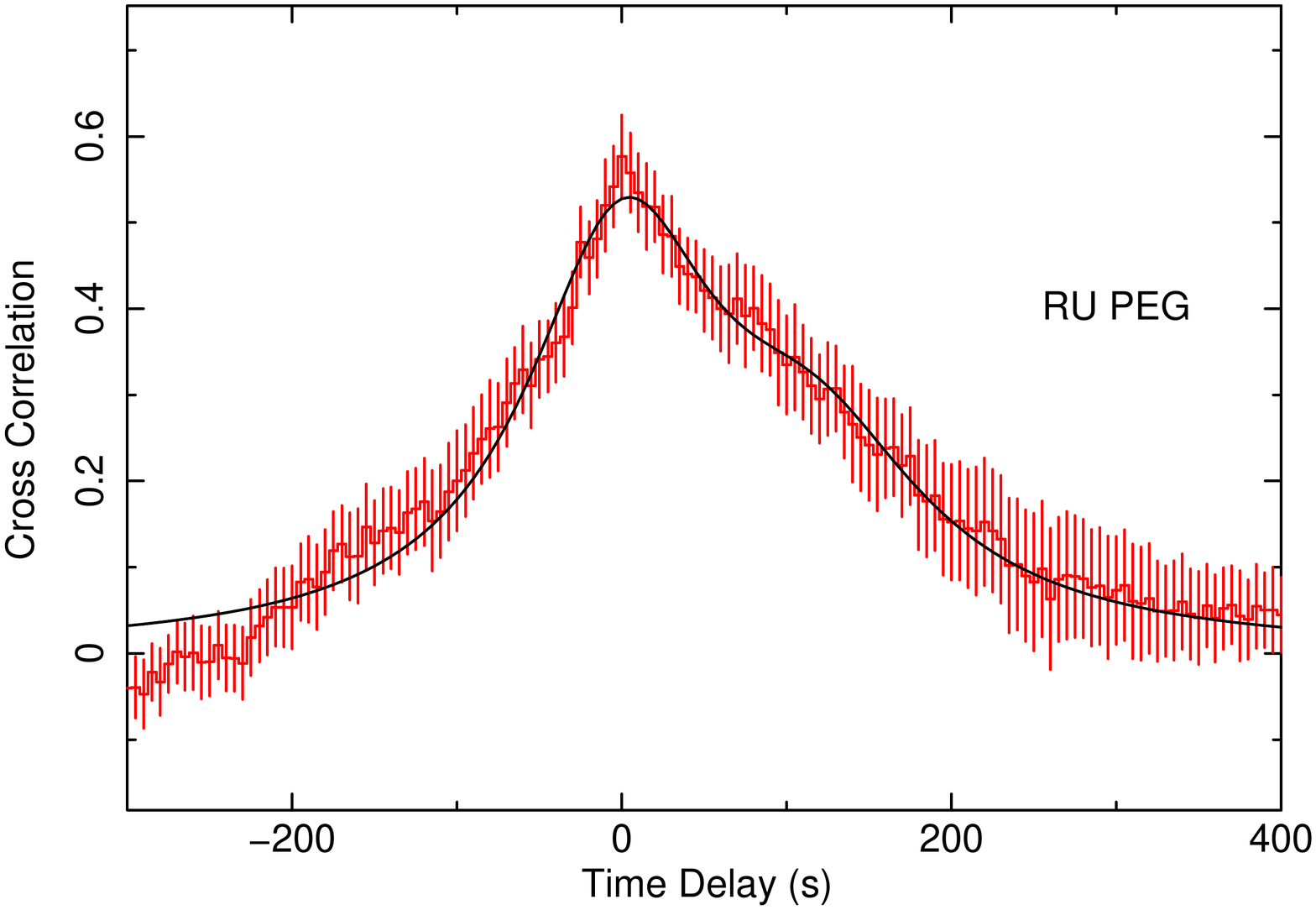}}
\centerline{
\includegraphics[width=8.1cm,height=4.3cm,angle=0]{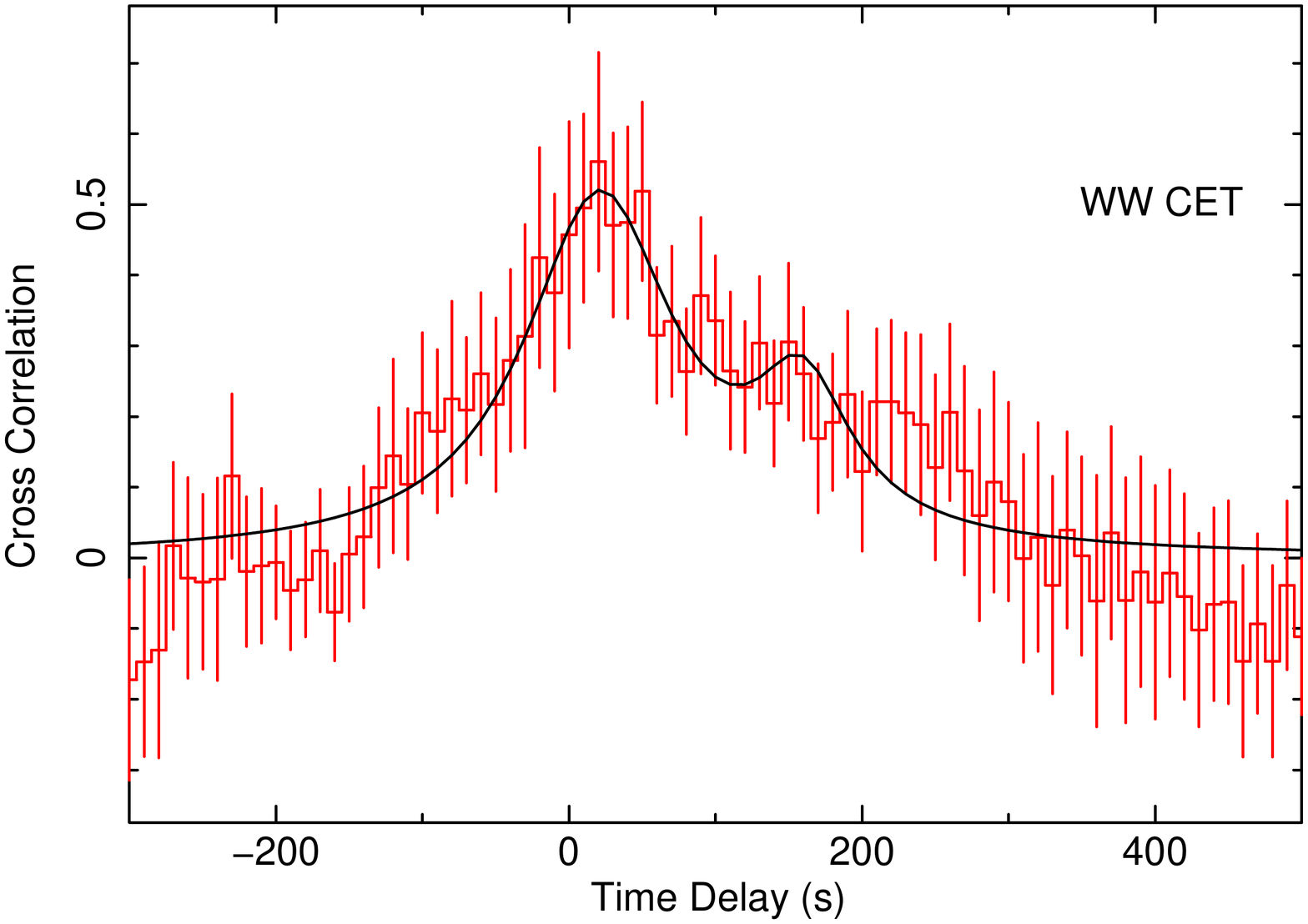}
\includegraphics[width=4.3cm,height=8.5cm,angle=90]{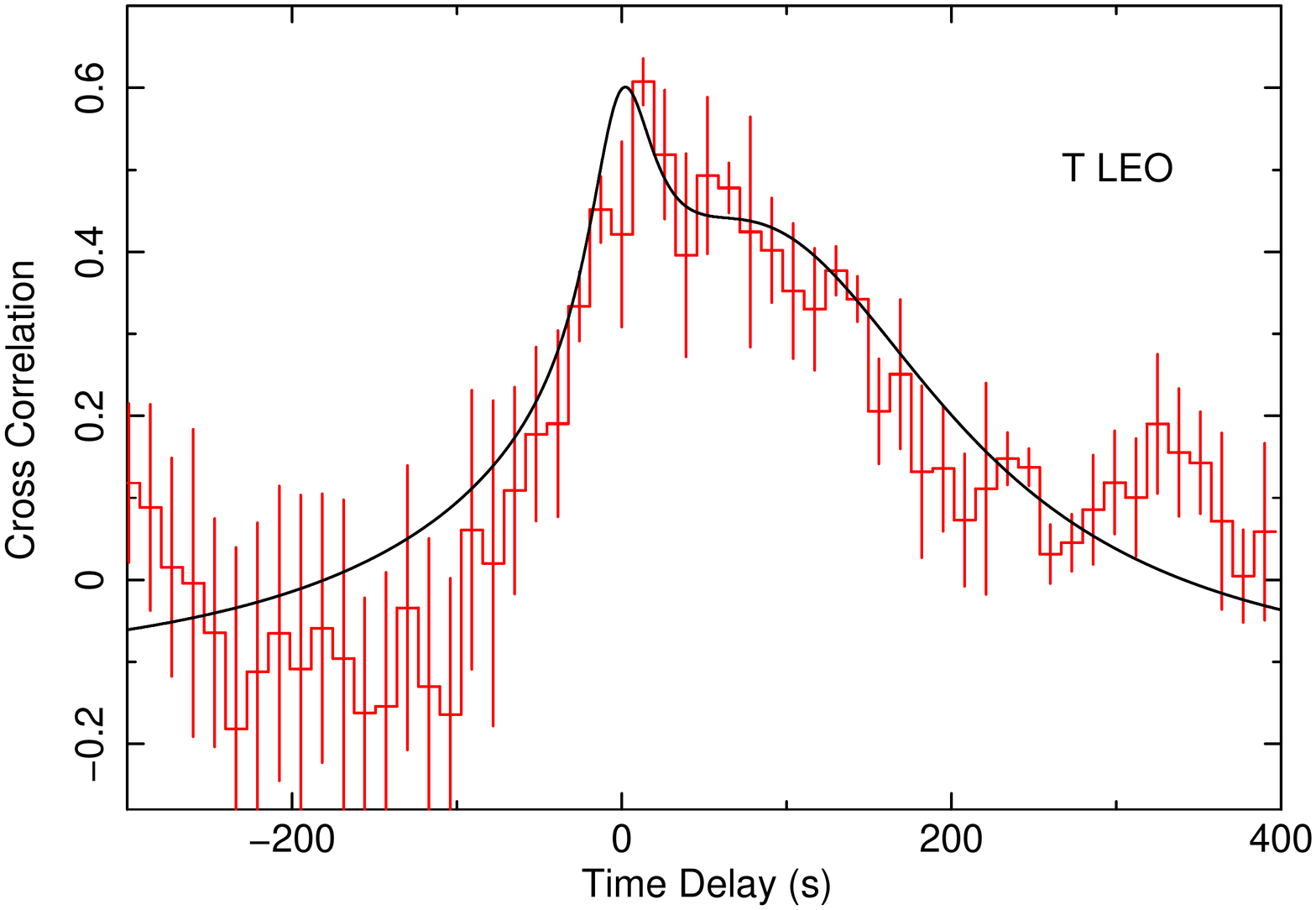}}
\centerline{
\includegraphics[width=8.1cm,height=4.3cm,angle=0]{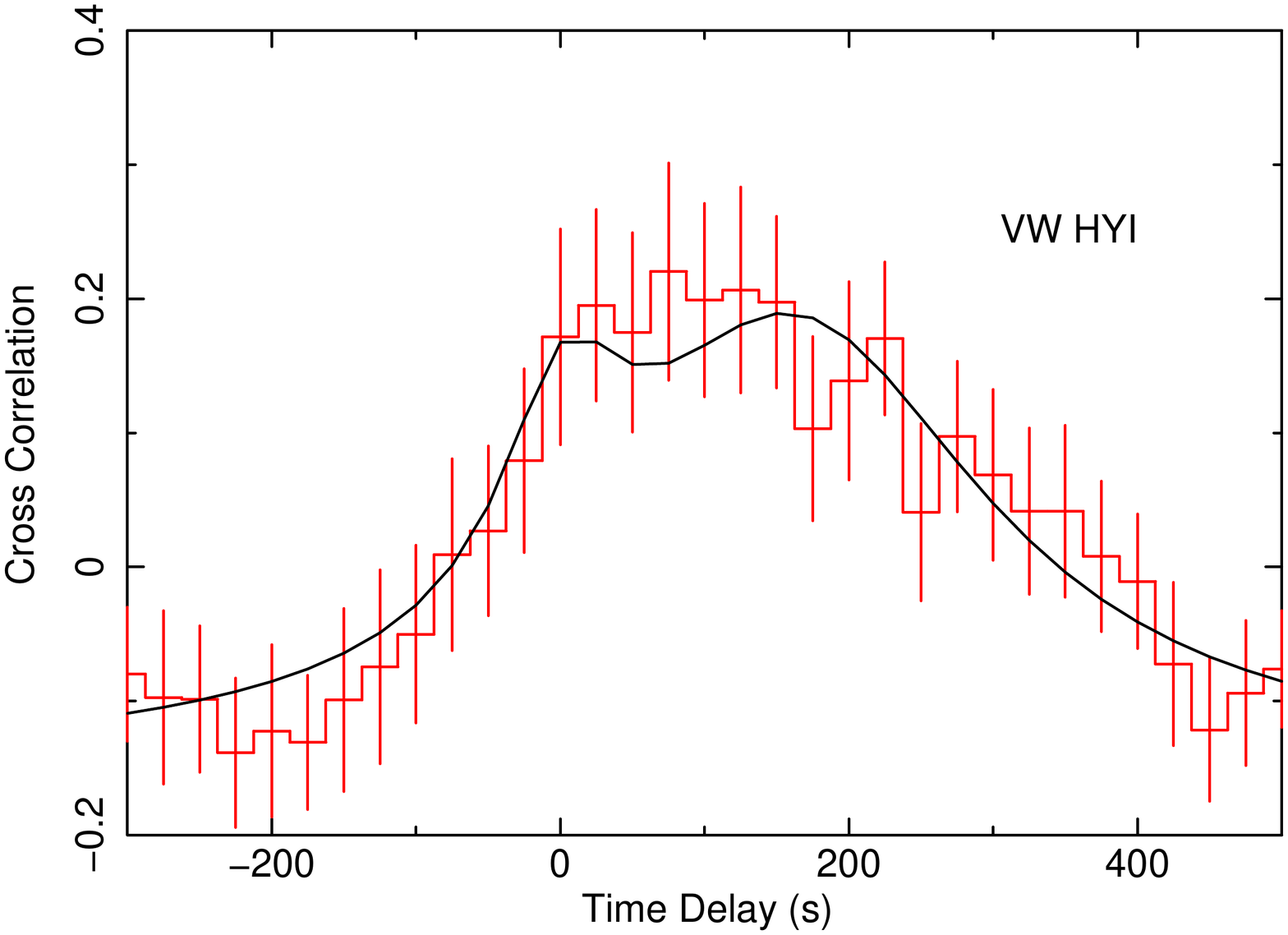}
\includegraphics[width=8.1cm,height=4.3cm,angle=0]{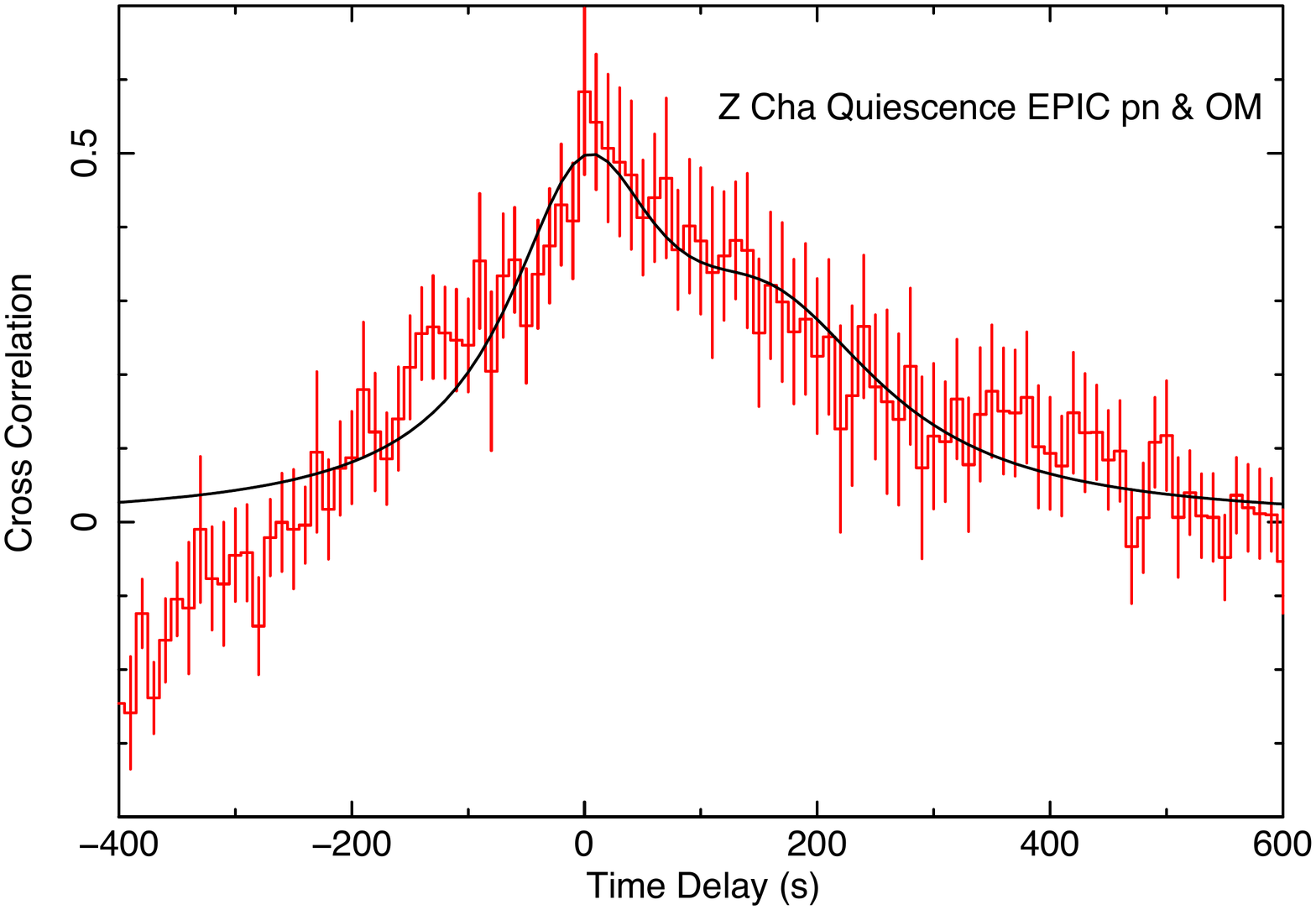}}
\caption{The cross-correlation of the EPIC pn (X-ray) and OM (UV) light curves with 1 sec time resolution.  The
CCFs are displayed for SS Cyg, RU Peg, WW Cet, T Leo, VW Hyi and Z Cha from the top left to the bottom right of the figure.
The correlation coefficient is normalized to a maximum value of 1. The two-component Lorenzian fits
are shown as solid black lines (except for SS Cyg where a single Lorentzian was used). See \citet{2012Balman}.}
\end{figure*}

\begin{table}
\label{1}
\caption{The break frequencies, disk transition radii, and time lags/delays of the dwarf novae \citep{2012Balman,2019Balman}}
\begin{center}
\begin{tabular}{ccc}
\hline
\hline

\multicolumn{1}{c}{Source} &
\multicolumn{1}{c}{Break Freq. (mHz)} &
\multicolumn{1}{c}{Delay (s)} \\
\hline

SS Cyg (quies) & 5.6$\pm$1.4 &  166-181 \\
SS Cyg (X-ray Dips)  & 50.0$\pm$20.0 &  N/A \\
SS Cyg  (X-ray peak) & 9.7$\pm$1.5 &   N/A  \\
RU Peg  (quies) &2.8$\pm$0.5  &  97-109 \\
VW Hyi  (quies) & 2.0$\pm$0.6 &  103-165  \\
WW Cet  (quies) & 3.0$\pm$1.7 & 118-136  \\
T Leo  (quies)  & 4.5$\pm$1.5 &  96-121  \\
HT Cas  (quies)  & 4.8$\pm$2.4 &  N/A  \\
WZ Sge  (quies)  & 1.1$\pm$0.5 &  N/A  \\
WZ Sge (outb peak)  & 300  &  N/A \\
V426 Oph (quies)  & 3.6$\pm$1.0 &  182-245 \\
Z Cha  (quies)  & 1.0$\pm$0.4 &  112-209   \\
SU UMa (quies) & 6.5$\pm$1.0 & N/A \\
SU UMa (X-ray Dips) &  100 &  N/A \\

\hline
\end{tabular}
\end{center}
{\bf Note.}
The errors represent 90 $\%$ confidence level.
\end{table}

\section{The outflows of high state cataclysmic variables (nova-likes) and dwarf novae}

For accretion rates $\dot M_{acc}$$\ge$10$^{-9}$ M$_{\odot}$ yr$^{-1}$ in high state CVs (NLs) winds are readily observed  with wind mass loss rates at an efficiency
of $\le$ 1\%  of  the accretion rate. The outflow velocities are 200-5000 km/s  \citep{2002Long,2004Kafka,2011Puebla}. 
Several divisions exists within this group: VY Scl-types  (show occasional low states) and UX UMa-types (are always at high state)  show  emission lines, UX UMa 
stars show broad absorption features in the optical and/or UV spectra, and RW Tri stars are eclipsing UX UMa systems  (Warner 
1995).  In addition, SW Sex stars are a specific spectroscopic class  with P$_{orb}$=3-4 hrs \citep{2007Rodriguez-Gil}.
All NLs  reside above the period gap (concentration P$_{orb}$=3-4 hrs)  except BK Lyn below the period gap.     
Bipolar or rotationally symmetric winds and outflows are characteristic of NLs. Such wind outflows are best detected in the FUV with the P Cygni Profiles of  the resonant 
doublet CIV ${\lambda}$1549.5 (also detected with Si IV ${\lambda}$1397.6, N V ${\lambda}$1240.8) \citep{1982Guinan,1985Sion}.
DNe are known to show similar wind outflows during the outburst stage as they phase into the high state. There is only one DN detected with a wind in the quiescent stage, WX Hyi \citep{2003Perna}.
The wind outflows are modeled through hydrodynamical radiation transfer codes as radiative winds or line driven winds  \citep{1998Proga,1999Knigge,2000Drew}.
Though mechanisms of acceleration of the observed winds are still unclear, studies of winds in eruptions of DNe have focused on line-driven winds \citep{1999Feldmeier,2006Pereyra}.

\begin{figure*}\label{fig:8}
\centerline{
\includegraphics[width=8.5cm,height=7cm,angle=0]{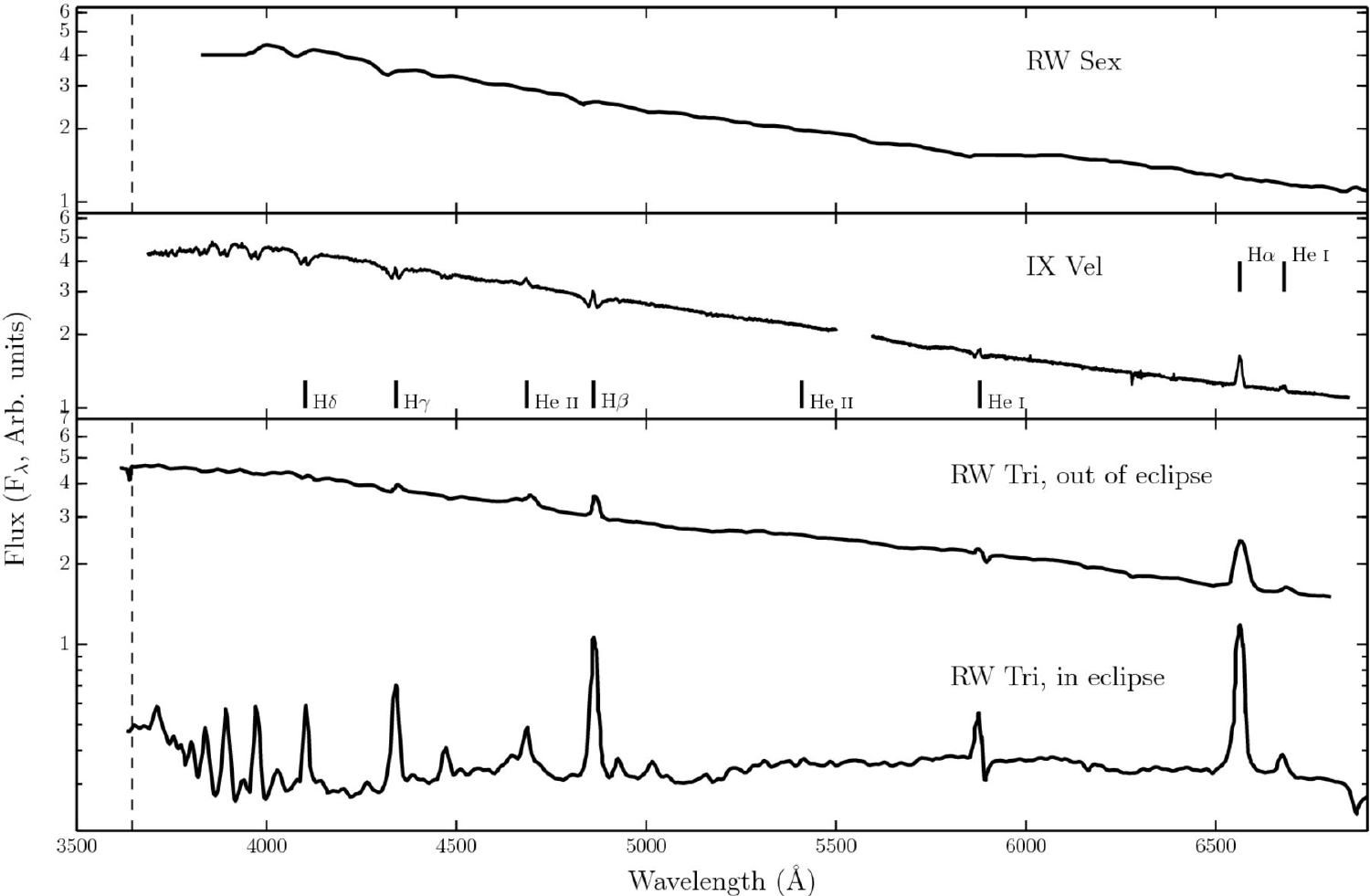}
\includegraphics[width=8.5cm,height=7cm,angle=0]{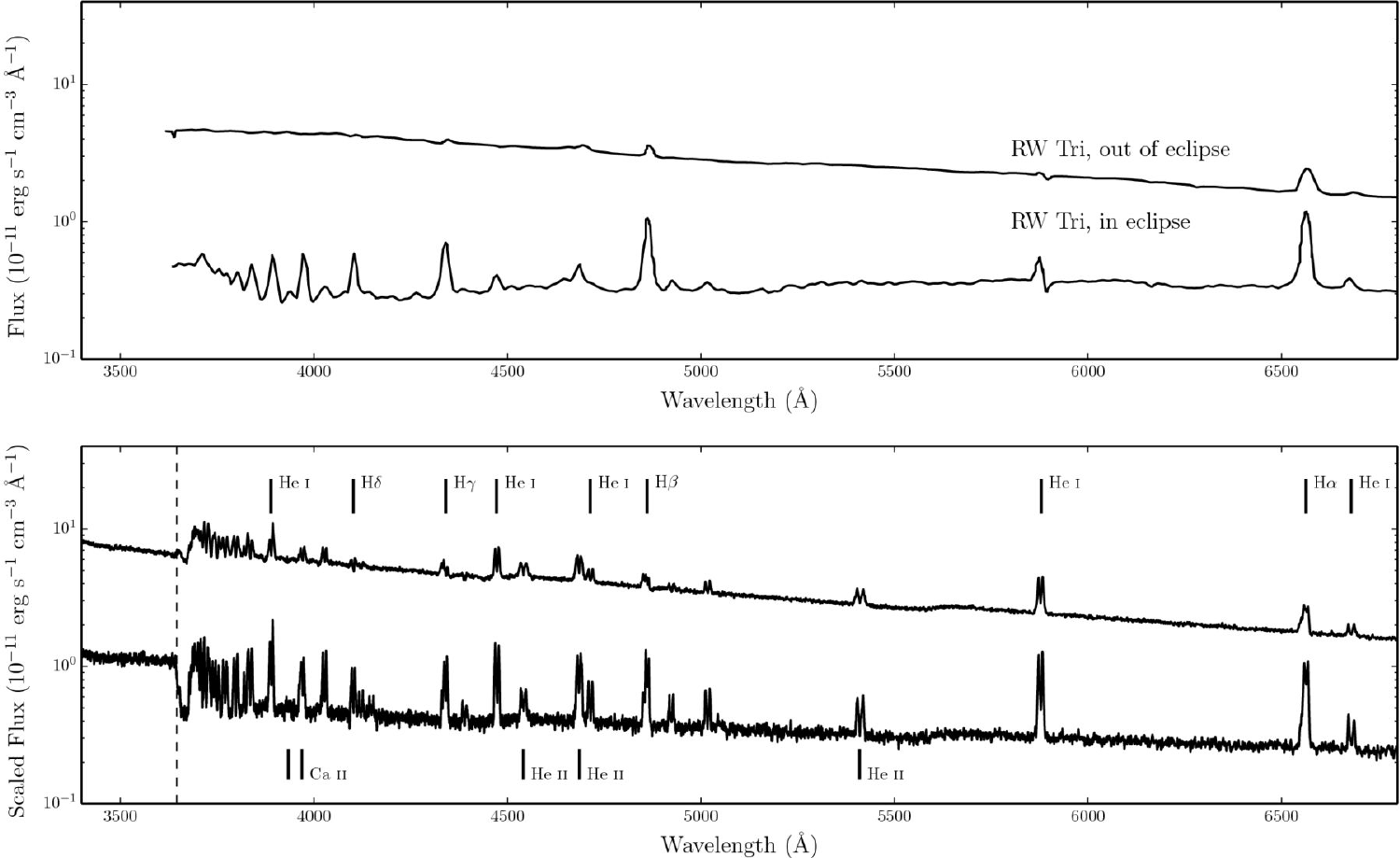}}
\caption{The left hand panels show the optical spectra of three NLs. On the right hand side; top Panel is the in and out of eclipse spectra 
of the high inclination NL RW Tri and bottom panel shows the in and out of eclipse, synthetic spectra. Figure is obtained from \citet{2015Matthews}.}
\end{figure*}

In many cases of NLs, and particularly in the SW Sex subclass of NLs, emission lines are single-peaked
\citep{1986Honeycutt,1995Dhillon,2004Groot}. 
However, theoretical expectations for lines formed in accretion disks are predicted to be double-peaked \citep{1981Smak,1986Horne}.  The expected dependence of this
on the inclination is not observed for NLs indicative of extended structures.
Low-state CVs (dwarf novae in quiescence) exhibit such double-peaked lines \citep{1990Marsh}.
Attempts to fit the observed spectral energy distributions (SEDs) of high-state CVs with models have not been as successful. 
In particular, the SEDs predicted by most stellar/disk atmosphere models are too blue in the UV \citep{1988Wade,1994Long,1998Knigge} and exhibit 
stronger-than-observed Balmer jumps in absorption \citep[][and references therein]{1998Knigge,2015Matthews}. In order to account for the discrepancies 
Monte Carlo radiative transfer simulations 
have been conducted to  test if the disk winds developed to account for the UV resonance lines would also produce significant amounts of optical line and/or 
continuum emission to produce the observed spectra and have found that it does produce some of the features but not all characteristics can be 
accounted \citep{2015Matthews}.
These authors model the eclipsing NL RW Tri and can not reproduce the
single peaked emission lines in the optical spectra unless an extension of about 150R$_{wd}$ above the disk is introduced (and/or extreme densities in the wind). 
Figure 8 shows optical spectra of few different NLs on the left hand panel. In the same figure on the right, 
top panel shows the observed spectra of RW Tri and the bottom panel displays the
synthetic spectra produced with the Monte Carlo radiative transfer simulations \citep{2015Matthews}. 
 \citet{2011Puebla}  have developed a spectrum synthesis method with disk-atmosphere models fitting the UV spectra from NLs. They find that the wind temperature is affected
 by the accretion rate and the primary mass have a strong effect on the P-Cygni profiles. 
They calculate a rather lower extended wind region of 17-18R$_{wd}$ for RW Tri  (compared to Mathews et al. 2015) and their analysis show that in NLs, wind  
extensions (vertical) are at most of the order of disk (radial) size; generally about 10R$_{wd}$.

In nonmagnetic CVs wind lines are found modulated on the orbital period \citep[e.g., six NLs,  six DNs][]{2000Prinja,2004Prinja,2009Kafka}.
A complex mixture of high and low ionization state lines are detected in the wind emission which signifies that  possible condensed high density regions  exists within the flow
as seen in FUV and  UV wavelengths \citep{2004Long,2006Long,2010Noebauer,2011Puebla}. These winds indicate
no disk precession or superhump effects though they originate from the disk. Variability timescales in the outflows are minutes to 100 s. Not much line variability is 
detected below 100 s. 
The existence of single peaked lines in high inclination systems are seen in several systems which create problems in modeling of the outflows as mentioned above.
There is a major lack of correlation between wind activity and system luminosity \citep{2002Hartley,2012Froning}.  All these complexities suggest that the 
winds in CVs and accreting WDs can not just be modeled through radiative or line driven winds, but possibly magnetic fields needs to be
introduced with magnetohydrodynamical calculations (MHD)  in order to account for the vast inhomogenities, time-variability and orbital modulations as also detected in XRBs. 
Recent work by \citet{2018Scepi} about wind-driven transport, accretion and  the light curves of DNe show that MHD winds may be present when a 
magnetic field as low as few tens of Gauss threads the disk.  They perform 3D local MHD shearing-box simulations including vertical stratification, radiative transfer, and a net constant vertical magnetic flux to investigate how transport changes between the outburst and quiescent states of DNe. They find that a net vertical constant magnetic field, provided by the white dwarf or by its stellar companion, provides a higher $\alpha$ of $\ge$0.1 in quiescence than in outburst, as opposed to what is expected. The derived winds are very efficient in removing angular momentum but do not heat the disk and enhances the accretion of matter, resulting in light curves that look like DNe outbursts. Other solutions are possible where the mass accretion rate is high but the density is lower than in a standard accretion disk, similar to the jet-emitting disk solutions.

\section{X-ray observations of classical and recurrent novae}

Classical novae (CNe) outbursts occur in AWB systems on the surface of the WD as a result of an explosive burning of
accreted material (Thermonuclear Runaway-TNR) causing the ejection of 10$^{-7}$ to 10$^{-3}$ M$_{\odot}$ of material at velocities
up to several thousand kilometers per second \citep{1989Shara,1994Livio,2001Starrfield,2008Bode}.
The classical nova systems have an initial low level accretion phase
($\le$10$^{-10}$ M$_{\odot}$ yr$^{-1}$; see Hillman et al. in this Special Issue), where the recurrent nova (RN) that is found in outburst 
with 20-100 years of occurance time, generally show
higher accretion rates at the level of 10$^{-8}$ -10$^{-7}$ M$_{\odot}$ yr$^{-1}$  \citep{2013Anupama}. Once the critical pressure at the base of the
WD envelope is reached (e.g., 10$^{19}$ dyn cm$^{-2}$) the temperatures reach T$\sim$10$^8$ K under semi-degenerate conditions, and a 
TNR occurs, resulting in the explosive burning of the hydrogen on the surface of the WD. The nuclear reactions produce 
Beta-unstable nuclei (isotopes) through the inverse Beta-decay
that reach the outer layers since the envelope is convective once the burning starts. These isotopes
dump enough energy to cause ejection of the envelope matter. A gradual spectral hardening of the
stellar remnant WD spectrum with time past visual maximum is expected to occur required by the  H-burning phase at
constant bolometric luminosity where 
photospheric radius of the WD decreases and the envelope mass is lost in outflows.
The residual hydrogen-rich envelope matter is consumed by H-burning and wind-driven
mass loss \citep{1995Prialnik,2005Yaron,2013Jose,2014Kato,2016Starrfield}.

The emission from the remnant WD is a blackbody-like stellar continuum
referred to as  the supersoft X-ray emission component.
As the stellar photospheric radius decreases in time during the constant bolometric
luminosity phase, the
effective photospheric temperature increases
(up to values in the range 1--10 $\times 10^5$ K) and the peak of
the stellar WD spectrum is shifted from the visual through the ultraviolet to the soft X-rays
(0.1-1.0 keV), where finally the H-burning turns off  
\citep{1998Balman,2001Balman,2002Orio,2007Ness,2010Rauch,2011Schwarz,2015Osborne,2017Balman,2018Orio}.
In addition, super soft X-ray emission from a polar cap of a magnetic  WD can yield a blackbody-like spectrum during the outburst stage
as a result of onset of mass accretion \citep[e.g., see][]{2017Balman}.

CNe, also,  emit hard X-rays (above 0.5 keV) as a
result of shocked plasma emission 
during the outburst stage referred as the hard X-ray emission component.
The main mechanisms responsible for this component are : (1)
circumstellar interaction
\citep{2005Balman,2014Balman-aa,2006Bode,2006Sokoloski,2009Takei,2016Cheung}; X-rays can be thermal or non-thermal in origin
(2)  wind-wind interaction
\citep{2001Mukai,2005Orio,2008Lynch,2009Ness,2014Chomiuk}; (3) stellar wind instabilities and X-ray
emission \citep[as in][]{1999Owocki};  and (4) mass accretion, detection of 
flickering in the X-ray light curves or detection of the spin period, and detection of the 6.4  keV  fluorescence Fe line
\citep[e.g.,][]{2002Hernanz,2007Hernanz,2010Page,2015Osborne,2018Aydi}.
Comptonized X-ray emission from the Gamma-rays produced in radioactive decays
after the TNR has been suggested as a possible hard X-ray Component 
\citep[example: 22Na, 7Be, 26Al;][]{2002Hernanz-nar,2012Hernanz}, but never observed. Gamma-rays from novae during eruption has been observed 
with Fermi \citep[e.g.,][]{2010Abdo,2014Ackermann} , and  the origin of this emission is believed to be the particle acceleration in shocks of the ejected material
in the polar and equatorial regions (from different winds/jets) around the nova \citep{2014Chomiuk}. Gamma-ray emission from the shocks in the circumstellar medium of
novae during outburst opens a new perspective to nova explosions and its energetics. More recently, simultaneous detection of Gamma-ray emission with Fermi and hard 
X-ray emission ($>$10 keV)  using  \nustar\ were made (V5855 Sgr) that show the relation of the Gamma-rays from particle acceleration and thermal X-rays in the shock emission \citep{2019Nelson}. 
 A new finding reports simultaneous optical and Gamma-ray observations of distinct correlated series of flares in both bands for a nova that exploded in 2018; V906 Car \citep{2020Aydi}. 
 The optical flares lag the Gamma-ray flares slightly, suggesting origins in radiative optical shocks. During the flares the nova luminosity doubles showing for the first time the bulk of luminosity of a nova can be shock-powered. 
The ejected matter  from novae appear to consist of two basic main components: a slow, dense outflow with a maximum velocity of 
less than about 1000 km s$^{-1}$ and a fast outflow or wind with a maximum expansion speed of several thousand km s$^{-1}$ and as the 
nova remnant expands, the slow flow is observed as a dense ring/core in radio or optical images, whereas the fast flow appears as more extended, 
bipolar lobes (several outflows can be detected) \citep{2017Mukai,2017Sokoloski}. 
The X-ray remnant shells/ejecta are few (two-three) and can only be resolved and detected years later \citep{2005Balman,2014Balman-aa}.

As a result, accretion is the cause of these TNR events (novae phenomenon) and novae have been detected as accreting sources even 
during the outburst phases as noted above. 
It is unclear how early the  accretion may be onset during the outburst stage but indications are as early as 40-50 days to 168 days  or 
further into two years after eruption \citep[e.g.,][and references therein]{2018Aydi,2017Balman}. Several old novae have been detected to 
resume accretion long after the outburst stage as a typical nonmagnetic CV, and MCV (or symbiotic system for that matter) 
with the characteristics discussed in the previous sections  \citep[see][]{2001Orio,2018Zemko}.  Furthermore, it has also been suggested for an
RN (i.e., T Pyx) that the TNR eruption and H-burning is initiated as a result of the advective heating of the WD from the advective hot flows 
 in the inner regions of a warped disk structure at high mass accretion rates typical for RN \citep{2014Balman-aa}.

\section{On the accreting white dwarf binaries that are related to cataclysmic variables}
\subsection{AM CVn systems}

AM CVn systems are a class of ultracompact binary systems with helium dominated spectra 
depicting 5-65 min orbital periods shorter than the period minimum for CVs \citep[see][for a review]{2005Nelemans,2010Solheim}. The prototype itself (AM CVn)
was discovered in 1967 with an orbital period of 18 min \citep{1967Smak}. 
These objects have evolved from one of the following scenarios:  i)  A double WD star system that first evolves to shorter periods due to 
angular momentum loss caused by  gravitational wave radiation, starting mass transfer at orbital periods of a few minutes, and 
then evolving to longer periods with decreasing mass transfer rate;  
ii) a WD-low mass and a non-degenerate helium star binary that transfer mass as it evolves to a minimum period of 10 min when the star
becomes semi-degenerate; and iii) as CVs with evolved secondaries that lose their outer hydrogen envelope, uncovering their He-rich 
core, and then evolve as helium stars.  

These objects show variability on several timescales dependent on orbital period and mass 
transfer rate in the systems. As a function of increasing orbital period and time as they evolve to longer periods they go through three 
distinct phases. A high state phase with systems at P$\le$ 20 min corresponding to high accretion rates. The disk is optically thick  
and shows low amplitude periodicities at the orbital period and superhump period and/or their beat period \citep{2002Patterson,2005Patterson}. 
The latter occurs as the disk becomes 
eccentric and starts to precess as the result of extreme mass ratio \citep{1988Whitehurst}. Another one is the quiescent state associated with
low accretion rates and optically thin disk. These systems are associated with systems P $\ge$ 40 min. 
In this state not much optical photometric variability exists and thus systems are studied spectroscopically with only some showing
He absorption lines \citep{2005Roelofs}. A dominant phase is an outbursting state in which optical variability is seen with periods 
between 20 and 40 min. During these phases systems resemble the high-state and show absorption lines. Emission lines are visible in the
quiescent low mass transfer rate states \citep{2001Groot}. In outburst, these systems are 3-5 mag brighter than in quiescence with 
recurrence on timescales from 40 d to several years. Some systems, at short-periods, are also observed to have shorter, 
normal outbursts lasting 1-1.5 d seen 3-4 times between superoutbursts \citep[e.g.,][]{2011Levitan}.

A significant number of AM CVn systems have been found using photometric variability with large-area synoptic surveys: (1)
Palomar Transient Factory (PTF)  \citep{2015Levitan}; (2) Catalina Real-Time Transient Survey (CRTS) \citep{2014Breedt}. 
The PTF search for AM CVn systems has provided identified systems without the use of color ,
in order to test the  current population models. However, this needs a well developed model for their outburst 
classification which is described in the framework of DIM used for DNe
\citep{1997Tsugawa,2012Kotko,2016Hameury},
although the changes in outburst patterns for AM CVn systems \citep[e.g., CR Boo][]{2000Kato}  are not well explained.
The observed space density of AM CVn systems is 3$\times$10$^{-6}$ pc$^{-3}$ \citep{2007Roelofs}, however, the modeled density is about a factor of 10
larger.  $Gaia$ DR2 data are revising our view of these systems; for example a collection of 15 AM CVns observed with $Gaia$ yield an accretion rate  range 
of 1.1$\times$10$^{-11}$--2.5$\times$10$^{-8}$ M$_{\odot}$ yr$^{-1}$ in the optical band that is significantly higher than what standard models predict \citep{2018Ramsay}.    

AM CVn stars have been detected as weak X-ray emitters \citep{1995Ulla,2006Ramsay,2007Ramsay}. The X-rays are suggested to come from 
the accretion disk and/or the wind and show no magnetic behaviour  \citep{1996vanTeeseling,2005Ramsay,2006Ramsay}. Even though
they show no coherent or quasi periodic pulsations in the X-rays, they show  orbital modulations and some suppression of X-ray emission 
during the outbursts \citep{1994vanTeeseling,2007Ramsay,2012Ramsay} which are characteristics similar to DNe (i.e., CVs).
The X-ray spectra are modeled best by thin thermal plasma with highly nonsolar abundances that
shows hydrogen deficiency and significant nitrogen over abundance indicative of nova eruptions. The X-ray plasma temperatures are cooler than DN in a range 2-8.4 keV
\citep{2005Ramsay,2012Ramsay}. The temperatures in quiescence and off-outburst 
show a power law distribution in the plasma with ${\alpha}$ =0.8-1.05, typical of DN
in quiescence. The spectra show
some X-ray suppression in flux and luminosity and cooling in hard X-ray temperatures during outburst (like DNe) , 
but by a factor of 1.5-2 \citep[][and references therein]{2012Ramsay,2019RiveraSandoval}. 
No soft X-ray emission as to a blackbody is detected from these systems on or off outburst (which is a characteristic of advective hot flows in the inner disk see sec. [2.3]). The X-ray luminosity of AM CVn systems 
are in a range 1.6$\times$10$^{33}$-1$\times$10$^{30}$ erg s$^{-1}$ but largely found to be 
L$_x <$10$^{32}$ erg s$^{-1}$, in general similar with DNe and NLs \citep{2005Ramsay,2006Ramsay}.
The UV luminosity of these systems show strong dependence on the 
orbital period by a change of a factor of 1500 for different orbital periods due to hotter disks or higher accretion rates at shorter periods 
whereas the X-rays show no such dependence (true for most CVs, as well) \citep{2007Ramsay,2010Solheim}.    
There are two interesting systems HM Cnc and V407 Vul  characterized by ‘on/off’ X-ray light curves that trail the optical light curves by  0.2 in phase 
\citep[][and references therein]{2004Cropper,2007Barros}. These systems show only soft X-rays and no hard X-rays, and the observed orbital periods are decreasing with time at a rate consistent with gravitational radiation  and
angular momentum loss. Flickering is minimal or absent.  The proposed model for these systems is the direct impact model with no disk involved \citep{2001Nelemans,2002Marsh}. At short enough orbital periods the accretion stream can impact the surface of the WD directly. The accreted matter would thermalize below the photosphere with a temperature cool enough to emit soft X-rays.

\subsection{White dwarf  symbiotics}

Symbiotic systems are interacting binaries where a hot compact primary star accretes matter from an evolved giant secondary star.
The primaries are mostly WDs but it can also be a neutron star.  In general, the primary underfills its Roche Lobe  with binary separations of 10$^{13}$-10$^{15}$ cm
and the accretion onto the primary  WD occurs through a stellar wind \citep{1944Bondi}. 
The angular momentum captured by the Bondi-Hoyle process allows for  
formation of accretion disks leading to the wind Roche-lobe overflow scenario \citep{1984Livio,2007Podsiadlowski,2011Alexander}. 
\citet{2007Podsiadlowski} proposed that  if the red giant does not fill its Roche lobe, its wind does and  the wind is collimated 
toward the L1 point of the orbit that may 
yield an accretion disk around the white dwarf.   Other accretion modes are proposed to overcome the small 
efficiency of wind Roche-lobe overflow as in \citet{2015Skopal}, where
the wind from the giant in some symbiotics can be focused by rotation of the giant, following the 
wind compression disk model from \citet{1993Bjorkman}. 
 
The binary system is engulfed by a dense nebula ionized by the UV radiation from the WD photosphere and/or the accretion disk. 
The orbital periods range from a few hundred to a few thousand days \citep{2000Belczynski}. 
WDs in symbiotics have masses of about 0.6M$_{\odot}$ \citep{2007Mikolajewska}, but masses close to the Chandrasekhar limit exist.  Near Chandrasekhar limit
 sources experience recurrent nova outbursts or produce strong, hard X-ray emission \citep{2006Sokoloski,2007Luna,2009Kennea,2018Luna-aa}.
Symbiotics are proposed and detected as progenitors for type-Ia supernovae either through single or double degenerate
 channels \citep{2010Wang,2010DiStefano,2012Dilday}. 

Symbiotic systems show several types of outbursts. First group is the slow novae that lasts decades as a result of TNRs on the surface of less 
massive WDs. If the WD is massive, this constitutes the second group of recurrent symbiotic novae which repeats every several tens of years. 
The third group is the classical symbiotic outburst \citep[e.g., Z And:][]{2006Sokoloski-apj}.  These systems experience
DN like outbursts that can be triggered as a result of the thermal-viscous instabilities or mass transfer enhancement  which may lead a TNR at 
times, creating the combination novae \citep{2018Bollimpalli}.
Thermal-viscous instabilities exist, even for large mass transfer rates, since critical accretion rate where disk is unstable is a strong function of 
disk radius \citep{2018Bollimpalli}.

About 25\% of the symbiotic systems  indicate X-ray emission which has been studied with several space observatories. 
The first survey has been with the \rosat\  
observatory \citep{1997Muerset,1996Muerset} and a detailed survey has been made including \swi\ \citep{2013Luna}.  The symbiotic systems can be 
categorized in three major groups with some showing the characteristics of both of the systems \citep{1997Muerset}. The ${\alpha}$ class is associated with quasi-steady 
H-burning (in a shell) below mostly 0.4 keV and nothing above 1 keV \citep{2007Orio}.   If a WD with shell burning is massive enough for 
photospheric temperatures to reach at least several hundred thousand degrees, burning symbiotics emit super soft X-ray emission 
\citep[e.g., AG Dra, RR Tel, and StH$\alpha$ 32][]{2017Sokoloski}. Such super soft X-rays (SSS, blackbody-like emission see sec. [5]) are detectable if the column of absorbing material is low. 
The ${\beta}$ class is the soft (-hard) X-ray emitters (i.e., plasma emission) below 2 keV originating in colliding winds in the systems.  The third group is the ${\delta}$ category 
where the X-ray emission is hard, highly absorbed and thermal above 2 keV.  The most overlap within these groups occur as 
the ${\beta}$/${\delta}$  type where there are two components of 
emission, showing both the soft and hard X-ray components. The hard absorbed components are believed to be originating in 
an optically thin boundary layer (BL).  The soft component originates in the colliding winds from the WD and the secondary giant star. 
It has also been suggested that if a Bondi-Hoyle accretion rate in the range 10$^{-8}$-10$^{-7}$ M$_{\odot}$
below 10$^{-7}$ M$_{\odot}$ is assumed, a 1.0-0.6 M$_{\odot}$ WD  will assume an optically thick BL, however the observed BLs have hard absorbed 
X-ray emission \citep{2013Luna} (i.e.,  similar problem to high state CVs as explained in sec. [2.2,2.3] using advective hot flows). For example, multi-satellite
X-ray observations of RT Cru indicate that the BL is optically thin and stays thin throughout a brightening event where the rate of accretion is 
increased to 6.7$\times$10$^{-9}$ M$_{\odot}$ yr$^{-1}$
\citep{2018Luna} (e.g., typically seen in DNe state changes). This observation also indicates a Compton reflection hump E$>$10 keV as detected in accreting (XRB) systems. 
In addition, the combined \swi, \nustar, and \suzaku\ analysis of T CrB indicates  that  in an active state that spans  few years, the system shows a 
hard X-ray emitting disk component that changes luminosities between several $\times$10$^{32}$-10$^{33}$ \lumcgs\ as in typical DN and 
only in the highest active state displays soft X-ray emission (a blackbody  with kT=35 eV) along with the hard X-ray component \citep[see][and references therein]{2019Luna}.
This system shows similarities to SS Cyg and indicates existence of advective hot flows in the inner disk as described in sec. (2\&3). During the most active phase of T CrB, where the soft X-rays are detected,
the hard X-rays shows the typical DN anticorrelation (seen in outbursts) where the temperature drops and the luminosity goes down by about a factor of 10. 

X-ray temporal characteristics lack coherent modulations as would be observed from accretion disks or WD spins, but detectable 
UV flickering is common in symbiotics \citep{2013Luna}. In WD symbiotics, only one system has been detected with a coherent modulation, 
as due to magnetic accretion, of optical emission with a  period of approximately 28 m \citep[Z And;][]{2006Sokoloski-apj}, in addition an oscillation 
with a period of 1734 s was marginally detected in X-rays from R Aqr \citep{2007Nichols}.
The UV variations are minutes to hours and support an accretion scenario rather than H-burning. 
The hard X-ray symbiotics show high UV and the soft X-ray 
symbiotics have low UV variability which support accretion scenario for the hard X-ray emitters, not the soft ones. In addition, 
the X-ray emission in the hard band have contribution from the jets detected in some of these systems. The jet (X-ray) luminosities 
are in a range  5$\times$10$^{30}$-5$\times$10$^{28}$ erg s$^{-1}$  \citep{2007Karovska,2007Nichols}  
which is only a small portion of the total X-ray luminosity that is  in a range 2.2$\times$10$^{33}$-3$\times$10$^{31}$ erg s$^{-1}$ . 
About 20\% of the symbiotics are associated with bipolar outflows \citep{1999Corradi}.  These jets and outflows are mostly linked with the
${\beta}$/${\delta}$  type symbiotics.  In the colliding winds scenario, the WD is assumed outbursting with winds and quiescent winds are 
rarely detected \citep[one system, EG And;][]{1993Vogel}. However, disk winds have not been ruled out.  Some accreting WDs  in symbiotics have 
been suggested to display  a softer spectrum from their boundary layers as a result of
Compton cooling at high accretion rates \citep{2011Nelson,2013Luna}.

\section{Final Remarks}

AWBs are important laboratories for accretion and outflow physics.  They, particularly CVs, have been known to host  standard, steady-state, constant 
rate accretion disk flows.  However, progress in multi-wavelength observations
in particular X-rays using moderate or high resolution spectral and timing data show that this knowledge is no longer reliable. The X-ray observations together
with other wavelength data indicate that accretion flows differ in the inner regions of the accretion disk (compared with the outer regions) and diverge from the standard
accretion disk picture  in quiescence (e.g., DNe) or high states (e.g., NLs and DNe) with extended regions (vertical and radial) particularly in the X-rays (and to some extend UV).  
MCVs with accretion disks (IPs) show accretion flow characteristics more inclined towards the standard disk scenario. The related symbiotic  and AM CVn systems indicate 
similar disk structure to nonmagnetic CV systems.
 
The X-ray broadband noise characteristics of at least 10 DNe in quiescence show frequency breaks at 
1-6 mHz, as a result of truncation of the optically thick accretion disk
at radii $\sim$(3.0-10.0)$\times$10$^9$\ due to the change in the  accretion flow characteristics. Furthermore,
these break frequencies are inversely correlated with the X-ray temperatures \citep[in quiescence and outburst][]{2019Balman}.
The observational characteristics of DNe  (particularly in the X-rays) can be readily explained 
with the existence of advective hot flows in  the X-ray emitting regions that are not removed in the outbursts.
In general, DNe broadband noise (quiescence and outburst) show close similarities to XRBs which indicates that
the accretion flows around a WD 
resemble to that of the black hole (BH) or neutron star (NS) accretors with an optically
thick colder outer accretion disk and an optically thin advective hot flow in the inner regions, e.g.,  close to the WD
\citep[see review for  XRBs by][]{2007Done}. The broadband noise in XRBs is also a power law with an $\alpha$ of 0.7-1.2 with sometimes
a break to an index about 1.5-2 at around 3000 mHz  which portrays a similar character to nonmagnetic CVs \citep[see comparative review by][]{2019Balman}.
In both NS and BH binaries during the state changes from low to intermediate states, all components of the
broadband noise increase in frequency become weaker and more coherent. Nonmagnetic CV PDS and broadband noise resembles PDS of 
high accretion rate states of BH/NS systems.The main PDS characteristics of nonmagnetic CVs indicate that regardless of the
type of DN (and perhaps high state CVs), X-ray broad-band noise in quiescence and its evolution in outburst  are similar, which is a result of the 
properties of the inner advective hot flow region. 

\citet{1993Narayan} show that the optically thin BLs of accreting WDs
in CVs can be radially extended and that they advect part of the
viscously dissipated energy as a result of their inability to cool, therefore
point out  that optically thin BLs can act as ADAF-like hot accretion flows \citep[see][for a more detailed discussion]{2014Balman}. 
However, no theoretical model has been attempted to derive characteristics of BL regions  that are merged with advective hot flows for WDs. 
The appearance of an advective hot flow (i.e., ADAF-like)
in the inner regions of the accretion disk will differ from
an ordinary rotating Keplerian disk  with larger $\alpha$ parameters 0.1-0.3 which will no longer be fully supported by rotation at sub-Keplerian speeds, 
with a significant radial velocity component.  Advection can play an important role for the energy budget and the emission
spectrum (i.e., X-rays) of the accretion flow with an inefficiency in radiation in the BL/inner disk regions that remains mostly optically thin (limited spectral features and  existence of nonequilibrium ionization plasma),
and heating of the central WD as 
part of the retained energy in the disk is transferred to the WD.  Radiative inefficiencies, 0.1-0.0001, of the X-ray luminosities (from optically thin X-ray regions)
are characteristics of nonmagnetic CVs. Similar luminosities and spectral characteristics (i.e., X-rays, in quiescence and outburst) are detected from related accreting WD systems which
imply similar accretion flow physics (for disks) that governs the AWBs in mostly the inner regions of the accretion disks (e.g., where X-rays are emitted and perhaps except for H-burning WDs).
Presently, disk models can accomodate disk truncation for optically thick flows even at high accretion rates, but inclusion of inner advective hot flow zones are yet to be explored, but vastly necessary
to compensate for physical presentation of accretion disks around (nonmagnetic) WDs and accretion history of these systems.  
Furthermore, the disk outburst mechanism, DIM and hysteresis calculations  for  the accretion disks around WDs need to
include the fact that  inner advective hot flows exists for a more realistic model that is in accordance with the X-ray and multi-wavelength observations. 

It is widely accepted that there is strong connection between ADAFs and outflows as in winds
and jets (collimated) outflows \citep{1995Narayan,1999Blandford}. 
This is largely because the ADAFs have positive Bernoulli parameter
defined as the sum of the kinetic energy, potential energy and enthalpy, thus the gas is not well bound
to the central star.  High state CVs and AWBs have winds with sometimes strong bipolar collimated outflows (or jets), at times
modulated with the orbital period  (e.g., CVs)
as detected in the optical, UV bands, and X-rays (e.g., symbiotics)  with velocities 200-5000 km s$^{-1}$. These are not well modeled in the present context (see sec. 4). 
CVs at high states have been
detected as nonthermal/thermal radio sources and symbiotics have jets that are nonthermal radio sources.  
Given these characteristics, the ADAF-like optically thin BLs or inner disk regions, may be the origin of
these outflows or may aid the formation of these outflows, winds, disk jets in AWBs in a way similar to XRBs or AGNs.  Outflows and jets from ADAFs are best studied  and modeled by
MHD calculations that are based on MRI-driven turbulence \citep{2014Yuan}, the recent work on the MHD based outflow models via MRI calculations for CVs (as discussed in sec. [4]) 
are promising approaches in this direction.
Note that in nova outbursts, the different kinds of winds and/or jet-like outflows arise predominantly from the WD (this is also true for symbiotics). In addition, super soft X-ray sources may thermally drive jets from their disks 
(at Eddington accretion rates).  

The ADAF-like  models in CVs/AWBs should utilize
reemission of the energy advected by the flow, modeled as a blackbody of temperature:
T$_{eff}=(L_{adv} / 4f\pi\sigma R_{wd}^2)^{1/4}$
where f is the fraction of the stellar surface that is emitting which is expected to be the WD surface (i.e., f=1.0).  \citet{2003Godon} 
have calculated that WDs in DN are not generally heated to more than 15\%  for cool WDs and to about 1\% for hot WDs above their original temperature
via standard disk accretion, even through multiple DN outbursts. Thus, advective heating can explain particularly high WD temperatures 
typical in NL systems and heating of WDs during DN outbursts (as a different accretional heating mechanism).  

Effects of advective hot flows (ADAF-like) on nova explosions will be dominantly in the evolution of the system into the nova outburst. As the WDs are heated by advective hot flows,
and at high accretion rate and accretion luminosity, more energy can be transferred (perhaps via weak shocks into the WD) and thus,  an instability-driven TNR explosion can take place in semi-degenerate
conditions on the surface of the WD. This was suggested for a system T Pyx  (a recurrent nova with a low WD mass not predicted by the standard nova theory for an RN). Such TNR explosions may not be occurring with regular time intervals while the
intervals of explosions may be increasing or decreasing in time depending on the accretion history of the system and the heating of the WD.  Note here that the advective heating (by accretion) may possibly produce instability-driven 
SN Type Ia explosions in WD binaries (in single degenerate systems) via built-in instabilities in the envelope without the need for Chandrasekhar limit for the mass of the WD. 
How this can be achieved and consequences of this is beyond the scope of this paper. 

Advective hot flows will have implications on CV evolution. As mentioned above, WDs will be hotter. This should increase the level of irradiation
of the secondary and possibly enhance the accretion flow from the secondary. However,  the gravitational energy release will decrease since the energy
is retained in the flow and not radiated (in the inner disk). This should slow down the evolution and/or stagnate it particularly above and below the period gap with a large group of sources stacked in the 3-4 hr range with hot WDs (i.e., as observed in NLs). For the sources below the gap since gravitational radiation is less, the shrinking 
of the orbit will also slow down while the secondary evolves in its Roche Lobe. As a result, the systems will reach the period minimum, at longer periods than theoretically expected which seems to be the case (see sec. 2). It is also possible that systems will not bounce to longer periods  from the period minimum which is never reached and thus create a CV graveyard that could explain the large lack in the post-period minimum systems (see sec.2). In addition, great care should be given to calculate space densities of CVs  (e.g., of nonmagnetic nature) from observations using particularly X-ray luminosity functions, as a result of inefficiency of emission from the inner advective hot flows. 
This can lead to underestimation of the population density.

The X-ray observations (of accretion) in nonmagnetic CVs and related systems suffers from sensitivity of the present X-ray telescope instrumentation since they are dim sources. Thus, detailed analysis regarding emission and absorption lines and their diagnostics, warm absorbers, and PDS analysis is at times not well sampled or statistically inadequate due to this problem. As a result, the some of the new or up-coming missions line \nustar, \textit{NICER}, \textit{eROSITA}, \textit{Athena}, and \textit{XRISM} which enable better/higher spectral resolution at higher sensitivity should teach us more on the accretion physics and the existing advective hot flows in AWBs, revealing their nature and characteristics to a greater extend.

\bibliographystyle{model2-names}


\end{document}